\documentclass[aps,eqsecnum,preprint,floats,epsf,epsfig,nofootinbib]{revtex4}
\usepackage{epsfig}

\begin{document}
\def\be{\begin{eqnarray}}
\def\en{\end{eqnarray}}
\def\non{\nonumber}
\def\la{\langle}
\def\ra{\rangle}
\def\a{{\cal A}}
\def\B{{\cal B}}
\def\c{{\cal C}}
\def\d{{\cal D}}
\def\e{{\cal E}}
\def\p{{\cal P}}
\def\t{{\cal T}}
\def\nc{N_c^{\rm eff}}
\def\CP{{\it CP}~}
\def\CPP{{\it CP}}
\def\vp{\varepsilon}
\def\drho{\bar\rho}
\def\deta{\bar\eta}
\def\vma{{_{V-A}}}
\def\vpa{{_{V+A}}}
\def\J{{J/\psi}}
\def\ov{\overline}
\def\Lqcd{{\Lambda_{\rm QCD}}}
\def\pr{{ Phys. Rev.}~}
\def\prl{{ Phys. Rev. Lett.}~}
\def\pl{{ Phys. Lett.}~}
\def\np{{ Nucl. Phys.}~}
\def\zp{{ Z. Phys.}~}
\def\lsim{ {\ \lower-1.2pt\vbox{\hbox{\rlap{$<$}\lower5pt\vbox{\hbox{$\sim$}
}}}\ } }
\def\gsim{ {\ \lower-1.2pt\vbox{\hbox{\rlap{$>$}\lower5pt\vbox{\hbox{$\sim$}
}}}\ } }

\font\el=cmbx10 scaled \magstep2{\obeylines \hfill July, 2007}

\vskip 1.5 cm

\centerline{\large\bf Charmless Three-body Decays of $B$ Mesons}
\bigskip
\centerline{\bf Hai-Yang Cheng$^{1}$, Chun-Khiang Chua$^{2}$ and
Amarjit Soni$^3$}
\medskip
\centerline{$^1$ Institute of Physics, Academia Sinica}
\centerline{Taipei, Taiwan 115, Republic of China}
\medskip
\centerline{$^2$ Department of Physics, Chung Yuan Christian
University} \centerline{Chung-Li, Taiwan 320, Republic of China}

\medskip
\centerline{$^3$ Physics Department, Brookhaven National
Laboratory} \centerline{Upton, New York 11973}
\medskip

\bigskip
\bigskip
\bigskip
\centerline{\bf Abstract}
\bigskip

\small

An exploratory study of charmless 3-body decays of $B$ mesons is
presented using a simple model based on the framework of the
factorization approach. The nonresonant contributions arising from
$B\to P_1P_2$ transitions are evaluated using heavy meson chiral
perturbation theory (HMChPT). The momentum dependence of
nonresonant amplitudes is assumed to be in the exponential form
$e^{-\alpha_{_{\rm NR}} p_B\cdot(p_i+p_j)}$ so that the HMChPT
results are recovered in the soft meson limit $p_i,~p_j\to 0$.  In
addition, we have identified another large source of the
nonresonant signal in the matrix elements of scalar densities,
e.g. $\la K\ov K|\bar ss|0\ra$, which can be constrained from the
decay $\ov B^0\to K_SK_SK_S$ or $B^-\to K^-K_SK_S$.  The
intermediate vector meson contributions to 3-body decays are
identified through the vector current, while the scalar meson
resonances  are mainly associated with the scalar density. Their
effects are described in terms of the Breit-Wigner formalism. Our
main results are: (i) All $KKK$ modes are dominated by the
nonresonant background. The predicted branching ratios of
$K^+K^-K_{S(L)}$, $K^+K^-K^-$ and $K^-K_SK_S$ modes are consistent
with the data within errors. (ii) Although the penguin-dominated
$B^0\to K^+K^-K_{S}$ decay is subject to a potentially significant
tree pollution, its effective $\sin 2\beta$ is very similar to
that of the $K_SK_SK_S$ mode. However, direct \CP asymmetry of the
former, being of order $-4\%$, is more prominent than the latter.
(iii) For $B\to K\pi\pi$ decays, we found sizable nonresonant
contributions in $K^-\pi^+\pi^-$ and $\ov K^0\pi^+\pi^-$ modes, in
agreement with the Belle measurements but larger than the BaBar
result. (iv) Time-dependent \CP asymmetries in $K_S\pi^0\pi^0$, a
purely $CP$-even state, and $K_S\pi^+\pi^-$, an admixture of
$CP$-even and $CP$-odd components, are studied. (v) The
$\pi^+\pi^-\pi^0$ mode is found to have a rate larger than
$\pi^+\pi^-\pi^-$ even though the former involves a $\pi^0$ in the
final state. They are both dominated by resonant $\rho$
contributions.  (vi)  We have computed the resonant contributions
to 3-body decays and determined the rates for the quasi-two-body
decays $B\to VP$ and $B\to SP$. The predicted $\rho\pi,~f_0(980)K$
and $f_0(980)\pi$ rates are in agreement with the data, while the
calculated $\phi K,~K^*\pi,~\rho K$ and $K_0^*(1430)\pi$ are in
general too small compared to experiment. (vii)  Sizable direct
\CP asymmetry is found in  $K^+K^-K^-$ and $K^+K^-\pi^-$ modes.

\pagebreak

\section{Introduction}

Three-body decays of heavy mesons are more complicated than the
two-body case as they receive resonant and nonresonant
contributions and involve 3-body matrix elements. The three-body
meson decays are generally dominated by intermediate vector and
scalar resonances, namely, they proceed via quasi-two-body decays
containing a resonance state and a pseudoscalar meson. The
analysis of these decays using the Dalitz plot technique enables
one to study the properties of various resonances.  The
nonresonant background is usually believed to be a small fraction
of the total 3-body decay rate. Experimentally, it is hard to
measure the direct 3-body decays as the interference between
nonresonant and quasi-two-body amplitudes makes it difficult to
disentangle these two distinct contributions and extract the
nonresonant one.

The Dalitz plot analysis of 3-body $B$ decays provides a nice
methodology for extracting information on the unitarity triangle
in the standard model (SM). For example, the Dalitz analysis
combined with isospin symmetry allows one to extract the angle
$\alpha$ from $B\to\rho\pi\to\pi\pi\pi$ \cite{Quinn}. Recently, a
method has been proposed in \cite{CPS} for determining CKM
parameters in 3-body decays $B\to K\pi\pi$ and $B_s\to K\pi\pi$.
This method was extended further in \cite{GPSZ} to $\Delta
I=1,I(K^*\pi)=1/2$ amplitudes in the above decays and to $I=1$
amplitudes in $B_s\to K^*\ov K$ and $B_s\to \ov K^*K$.

Nonresonant 3-body decays of charmed mesons have been measured in
several channels and the nonresonant signal in charm decays are
found to be less than 10\% \cite{PDG}. In the past few years, some
of the charmless $B$ to 3-body decay modes have also been measured
at $B$ factories and studied using the Dalitz plot analysis. The
measured fractions and the corresponding branching ratios of
nonresonant components for some of 3-body $B$ decay modes are
listed in Table \ref{tab:BRexpt}. We see that the nonresonant
3-body decays could play an essential role in $B$ decays. It is
now well established that the $B\to KKK$ modes are dominated by
the nonresonant background. For example, the nonresonant fraction
is about 90\% in $\ov B^0\to K^+K^-\ov K^0$ decay. While this is a
surprise in view of the rather small nonresonant contributions in
3-body charm decays, it is not entirely unexpected because the
energy release scale in weak $B$ decays is of order 5 GeV, whereas
the major resonances lie in the energy region of 0.77 to 1.6 GeV.
Consequently, it is likely that 3-body $B$ decays may receive
sizable nonresonant contributions. At any rate, it is important to
understand and identify the underlying mechanism for nonresonant
decays.

\begin{table}[t]
\caption{Branching ratios of various charmless three-body decays
of $B$ mesons. The fractions and the corresponding branching
ratios of nonresonant (NR) components are included whenever
available. The first and second entries are BaBar and Belle
results, respectively. } \label{tab:BRexpt}
\begin{ruledtabular}
\begin{tabular}{l c c c c}
 Decay & BR($10^{-6}$) &  BR$_{\rm NR}(10^{-6})$  & NR fraction(\%) & Ref. \\
\hline
 $B^- \to\pi^+\pi^-\pi^-$ &  $16.2\pm1.2\pm0.9$ &
$2.3\pm0.9\pm0.5<4.6$  & $13.6\pm6.1$ & \cite{BaBarpipipi} \\
 & -- & -- & -- \\
 $B^-\to K^-\pi^+\pi^-$ & $64.1\pm2.4\pm4.0$ &
 $2.87\pm0.65\pm0.43^{+0.63}_{-0.25}$ & $4.5\pm1.5$ & \cite{BaBarKpipi} \\
 & $48.8\pm1.1\pm3.6$ & $16.9\pm1.3\pm1.3^{+1.1}_{-0.9}$ & $34.0\pm2.2^{+2.1}_{-1.8}$ & \cite{BelleKpipi}\\
 $B^-\to K^+K^-K^-$ & $35.2\pm0.9\pm1.6$~\footnotemark[1] &
 $50\pm6\pm4$  & $141\pm16\pm9$ & \cite{BaBarKpKpKm} \\
 & $32.1\pm1.3\pm2.4$~\footnotemark[2] & $24.0\pm1.5\pm1.5$~\footnotemark[3]
 & $74.8\pm3.6$~\footnotemark[3] & \cite{BelleKpKpKm} \\
 $B^-\to K^-K_SK_S$ & $10.7\pm1.2\pm1.0$ & & & \cite{BaBarKmKsKs}  \\
 & $13.4\pm1.9\pm1.5$ & & & \cite{Belle2004} \\
 \hline
 $ \ov B^0\to \ov K^0\pi^+\pi^-$ & $43.0\pm2.3\pm2.3$ & & & \cite{BaBarK0pippim}\\
 & $47.5\pm2.4\pm3.7$ & $19.9\pm2.5\pm1.6^{+0.7}_{-1.2}$ & $41.9\pm5.1\pm0.6^{+1.4}_{-2.5}$ & \cite{BelleK0pipi} \\
 $\ov B^0\to K^-\pi^+\pi^0$ & $34.9\pm2.1\pm3.9$ & $<4.6$ & & \cite{BaBarKppimpi0} \\
 & $36.6^{+4.2}_{-4.3}\pm3.0$ & $5.7^{+2.7+0.5}_{-2.5-0.4}<9.4$ & & \cite{BelleKppimpi0} \\
 $\ov B^0\to K^+K^-\ov K^0$ & $23.8\pm2.0\pm1.6$ & $26.7\pm4.6$ & $112.0\pm14.9$ & \cite{BaBarKpKmK0} \\
 & $28.3\pm3.3\pm4.0$ & & & \cite{Belle2004} \\
 $\ov B^0\to K_SK_SK_S$ & $6.9^{+0.9}_{-0.8}\pm0.6$ & & & \cite{BaBarKsKsKs} \\
 & $4.2^{+1.6}_{-1.3}\pm0.8$ & & & \cite{Belle2004} \\
\end{tabular}
\end{ruledtabular}
 \footnotetext[1]{When the intrinsic charm contribution
is excluded, the charmless branching ratio will become
$(33.5\pm0.9\pm1.6)\times 10^{-6}$.}
 \footnotetext[2]{When the contribution from $B^+\to\chi_{c0}K^+$
is excluded, the charmless branching ratio will become
$(30.6\pm1.2\pm2.3)\times 10^{-6}$.}
 \footnotetext[3]{Belle found
two solutions for the fractions and branching ratios. We follow
Belle to use the large solution.}
\end{table}

The direct nonresonant three-body decays of mesons in general
receive two distinct contributions: one from the point-like weak
transition and the other from the pole diagrams that involve
three-point or four-point strong vertices. For $D$ decays,
attempts of applying the effective $SU(4)\times SU(4)$ chiral
Lagrangian to describe the $DP\to DP$ and $PP\to PP$ scattering at
energies $\sim m_D$ have been made by several authors
\cite{Singer,KP,Cheng86,CC90,Botella} to calculate the nonresonant
$D$ decays, though in principle it is not justified to employ the
SU(4) chiral symmetry. As shown in \cite{CC90,Botella}, the
predictions of the nonresonant decay rates in chiral perturbation
theory are in general too small when compared with experiment.
With the advent of heavy quark symmetry and its combination with
chiral symmetry \cite{Yan,Wise,Burdman}, the nonresonant $D$
decays can be studied reliably at least in the kinematic region
where the final pseuodscalar mesons are soft. Some of the direct
3-body $D$ decays were studied based on this approach
\cite{Zhang,Ivanov}.

For the case of $B$ mesons,  consider the three-body $B$ decay
$B\to P_1P_2P_3$. Under the factorization hypothesis, one of the
nonresonant contributions arises from the transitions $B\to P_1
P_2$. The nonresonant background in charmless three-body $B$
decays due to the transition $B\to P_1 P_2$ has been studied
extensively
\cite{Deshpande,Fajfer1,Fajfer2,Deandrea1,Deandrea,Fajfer3} based
on heavy meson chiral perturbation theory (HMChPT)
\cite{Yan,Wise,Burdman}. However, the predicted decay rates are,
in general, unexpectedly large. For example, the branching ratio
of the nonresonant decay $B^-\to \pi^+\pi^-\pi^-$ is predicted to
be of order $10^{-5}$ in \cite{Deshpande,Fajfer1}, which is too
large compared to the limit $4.6\times 10^{-6}$ set by BaBar
\cite{BaBarpipipi}.  Therefore, it is important to reexamine and
clarify the existing calculations.

The issue has to do with the applicability of HMChPT. In order to
apply this approach, two of the final-state pseudoscalars in $B\to
P_1P_2$ transition have to be soft. The momentum of the soft
pseudoscalar should be smaller than the chiral symmetry breaking
scale of order 1 GeV. For 3-body charmless $B$ decays, the
available phase space where chiral perturbation theory is
applicable is only a small fraction of the whole Dalitz plot.
Therefore, it is not justified to apply chiral and heavy quark
symmetries to a certain kinematic region and then generalize it to
the region beyond its validity. In this work we shall assume the
momentum dependence of nonresonant amplitudes in the exponential
form $e^{-\alpha_{_{\rm NR}} p_B\cdot(p_i+p_j)}$ so that the
HMChPT results are recovered in the soft meson limit $p_i,~p_j\to
0$. We shall see that the parameter $\alpha_{_{\rm NR}}$ can be
fixed from the tree-dominated decay $B^-\to \pi^+\pi^-\pi^-$.

However, the nonresonant background in $B\to P_1P_2$ transition
does not suffice to account for the experimental observation that
the penguin-dominated decay $B\to KKK$ is dominated by the
nonresonant contributions. This implies that the two-body matrix
element e.g. $\la K\ov K|\bar ss|0\ra$ induced by the scalar
density should have a large nonresonant component. In the absence
of first-principles calculation, we will use the $\ov B^0\to
K_SK_SK_S$ mode in conjunction with the  mass spectrum in $\ov
B^0\to K^+K^-\ov K^0$ to fix the nonresonant contribution to $\la
K\ov K|\bar ss|0\ra$.

In this work, we shall study the charmless 3-body decays of $B$
mesons using the factorization approach. Besides the nonresonant
background as discussed above, we will also study resonant
contributions to 3-body decays. Vector meson and scalar resonances
contribute to the two-body matrix elements $\la P_1P_2|V_\mu|0\ra$
and $\la P_1P_2|S|0\ra$, respectively. They can also contribute to
the three-body matrix element $\la P_1P_2|V_\mu-A_\mu|B\ra$.
Resonant effects are described in terms of the usual Breit-Wigner
formalism. In this manner we are able to figure out the relevant
resonances which contribute to the 3-body decays of interest and
compute the rates of $B\to VP$ and $B\to SP$. In conjunction with
the nonresonant contribution, we are ready to calculate the total
rates for three-body decays.

It should be stressed from the outset that in this work we take
the factorization approximation as a working hypothesis rather
than a first-principles starting point. If we start with theories
such as QCD factorization \cite{BBNS}, or pQCD \cite{Li} or
soft-collinear effective theory \cite{SCET}, then we can take
power corrections seriously and make an estimation. Since
factorization has not been proved for three-body B decays, we
shall work in the phenomenological factorization model rather than
in the established theories such as QCDF. That is, we start with
the simple idea of factorization and see if it works for
three-body decays, in the hope that it will provide a useful
zeroth step for others to try to improve.

The penguin-induced three-body decays $B^0\to K^+K^-K_S$ and
$K_SK_SK_S$ deserve special attention as the current measurements
of the deviation of $\sin 2\beta_{\rm eff}$ in $KKK$ modes from
$\sin 2 \beta_{J/\psi K_S}$ may indicate New Physics in $b\to s$
penguin-induced modes.  It is of great importance to examine and
estimate how much of the deviation of $\sin 2\beta_{\rm eff}$ is
allowed in the SM. Owing to the presence of color-allowed tree
contributions in $B^0\to K^+K^-K_{S}$, this mode is subject to a
potentially significant tree pollution and the deviation of the
mixing-induced \CP asymmetry from that measured in $B\to J/\psi
K_S$ could be as large as ${\cal O}(0.10)$. Since the tree
amplitude is tied to the nonresonant background, it is very
important to understand the nonresonant contributions in order to
have a reliable estimate of $\sin 2\beta_{\rm eff}$ in $KKK$
modes.

The layout of the present paper is as follows. In Sec. II we shall
apply the factorization approach to study $B^0\to K^+K^-K_S$ and
$K_SK_SK_S$ decays and discuss resonant and nonresonant
contributions. In order to set up the framework for calculations
we will discuss $B\to KKK$ modes in most details. We then turn to
$K\pi\pi$ modes in Sec. III. The tree-dominated modes $KK\pi$ in
Sec. IV, and $\pi\pi\pi$ in Sec. V. In Sec. VI, we determine the
rates for $B\to VP$ and $B\to SP$ and compare our results with the
approach of  QCD factorization. Sec. VII contains our conclusions.
The factorizable amplitudes of various $B\to P_1P_2P_3$ decays are
summarized in Appendix A. The relevant input parameters such as
decay constants, form factors, etc. are collected in Appendix B.

\section{$B\to KKK$ decays}

For 3-body $B$ decays, the $b\to sq\bar q$ penguin transitions
contribute to the final states with odd number of kaons, namely,
$KKK$ and $K\pi\pi$, while $b\to uq\bar q$ tree and $b\to dq\bar
q$ penguin transitions contribute to final states with even number
of kaons, e.g. $KK\pi$ and $\pi\pi\pi$. We shall first discuss the
$b\to s$ penguin dominated 3-body decays in details and then turn
to $b\to u$ tree dominated modes. For $B\to KKK$ modes,  we shall
first consider the neutral $B$ decays as they involve
mixing-induced \CP asymmetries.

\subsection{$\ov B^0\to KKK$ decays}

We consider the decay $\ov B^0\to K^+K^-\ov K^0$ as an
illustration. Under the factorization approach, the $\ov B {}^0\to
K^+ K^- \ov K {}^0$ decay amplitude consists of three distinct
factorizable terms: (i) the current-induced process with a meson
emission, $\la \ov B^0\to K^+\ov K^0\ra\times \la 0\to K^-\ra$,
(ii) the transition process,  $\la \ov B^0\to \ov K^0\ra\times \la
0\to K^+K^-\ra$, and (iii) the annihilation process $\la \ov
B^0\to 0\ra\times \la 0\to K^+K^-\ov K^0\ra$, where $\la A\to
B\ra$ denotes a $A\to B$ transition matrix element. In the
factorization approach, the matrix element of the $\ov B\to\ov
K\,\ov K K$ decay amplitude is given by
 \be \label{eq:factamp}
 \la \overline K\, \overline K K|{\cal H}_{\rm eff}|\ov B\ra
 =\frac{G_F}{\sqrt2}\sum_{p=u,c}\lambda_p^{(s)} \la\overline K\, \overline K K|T_p|\ov B\ra,
 \en
where $\lambda_p^{(s)}\equiv V_{pb} V^*_{ps}$ and the explicit
expression of $T_p$ in terms of four-quark operators is given in
Eq. (\ref{eq:Tp}). The factorizable $\ov B^0\to K^+K^-\ov K^0$
decay amplitude is given in Eq. (\ref{eq:AKpKmK0}). Note that the
OZI suppressed matrix element $\la K^+ K^-|(\bar d d)_{V-A}|0\ra$
is included in the factorizable amplitude since it could be
enhanced through the long-distance pole contributions via the
intermediate vector mesons such as $\rho^0$ and $\omega$.
Likewise, the OZI-suppressed matrix elements $\la K^+K^-|(\bar
db)_\vma|\ov B^0\ra$ and $\la K^+K^-|\bar d(1-\gamma_5)b|\ov
B^0\ra$ are included as they receive contributions from the scalar
resonances like $f_0(980)$.

For the current-induced process, the two-meson transition matrix
element $\la \ov K {}^0 K^+|(\bar u b)_{V-A}|\ov B {}^0\ra$ has
the general expression~\cite{LLW}
 \be
 \la \ov K {}^0 (p_1) K^+(p_2)|(\bar u b)_{V-A}|\ov B {}^0\ra
 &=&i r
 (p_B-p_1-p_2)_\mu+i\omega_+(p_2+p_1)_\mu+i\omega_-(p_2-p_1)_\mu
 \non\\
 &&+h\,\epsilon_{\mu\nu\alpha\beta}p_B^\nu (p_2+p_1)^\alpha
 (p_2-p_1)^\beta,
 \en
where $(\bar q_1q_2)_\vma\equiv \bar
q_1\gamma_\mu(1-\gamma_5)q_2$. This leads to
 \be \label{eq:AHMChPT}
 A_{\rm current-ind}^{\rm HMChPT} &\equiv&\la K^-(p_3)|(\bar s
 u)_{V-A}|0\ra \la\ov K {}^0 (p_1) K^+(p_2)|(\bar u b)_{V-A}|\ov B {}^0\ra \non\\
 &=& -\frac{f_K}{2}\left[2 m_3^2 r+(m_B^2-s_{12}-m_3^2) \omega_+
 +(s_{23}-s_{13}-m_2^2+m_1^2) \omega_-\right],
 \en
where $s_{ij}\equiv (p_i+p_j)^2$. To compute the form factors $r$,
$\omega_\pm$ and $h$, one needs to consider not only the
point-like contact diagram, Fig. 1(a), but also various pole
diagrams depicted in Fig. 1. In principle, one can apply HMChPT to
evaluate the form factors $r,~\omega_+$ and $\omega_-$ \cite{LLW}.
However, this will lead to too large decay rates in disagreement
with experiment \cite{Cheng:2002qu}. The heavy meson chiral
Lagrangian given in \cite{Yan,Wise,Burdman} is needed to compute
the strong $B^*BP$, $B^*B^*P$ and $BBPP$ vertices. The results for
the form factors are \cite{LLW,Fajfer1}
 \be \label{eq:r&omega}
 \omega_+ &=& -{g\over f_\pi^2}\,{f_{B_s^*}m_{B_s^*}\sqrt{m_Bm_{B_s^*}}\over
 s_{23}-m_{B_s^*}^2}\left[1-{(p_B-p_1)\cdot p_1\over
 m_{B_s^*}^2}\right]+{f_B\over 2f_\pi^2}, \non \\
 \omega_- &=& {g\over f_\pi^2}\,{f_{B_s^*}m_{B_s^*}\sqrt{m_Bm_{B_s^*}}\over
 s_{23}-m_{B_s^*}^2}\left[1+{(p_B-p_1)\cdot p_1\over
 m_{B_s^*}^2}\right], \non \\
 r &=& {f_B\over 2f_\pi^2}-{f_B\over
 f_\pi^2}\,{p_B\cdot(p_2-p_1)\over
 (p_B-p_1-p_2)^2-m_B^2}+{2gf_{B_s^*}\over f_\pi^2}\sqrt{m_B\over
 m_{B_s^*}}\,{(p_B-p_1)\cdot p_1\over s_{23}-m_{B_s^*}^2} \non \\
 &-& {4g^2f_B\over f_\pi^2}\,{m_Bm_{B_s^*}\over
 (p_B-p_1-p_2)^2-m_B^2}\,{p_1\!\cdot\!p_2-p_1\!\cdot\!(p_B-p_1)\,p_2\!\cdot\!
 (p_B-p_1)/m_{B_s^*}^2 \over s_{23}-m_{B_s^*}^2 },
 \en
where $f_\pi=132$ MeV, $g$ is a heavy-flavor independent strong
coupling which can be extracted from the CLEO measurement of the
$D^{*+}$ decay width, $|g|=0.59\pm0.01\pm0.07$ \cite{CLEOg}. We
shall follow \cite{Yan} to fix its sign to be negative. The
point-like diagram Fig. 1(a) characterized by the term
$f_B/(2f_\pi^2)$ contributes to the form factors $\omega_+$ and
$r$, while Figs. 1(b) and 1(d) contribute to $r$ and Fig. 1(c)
contributes to all the form factors.

\begin{figure}[t]
\vspace{-1cm}
  \psfig{figure=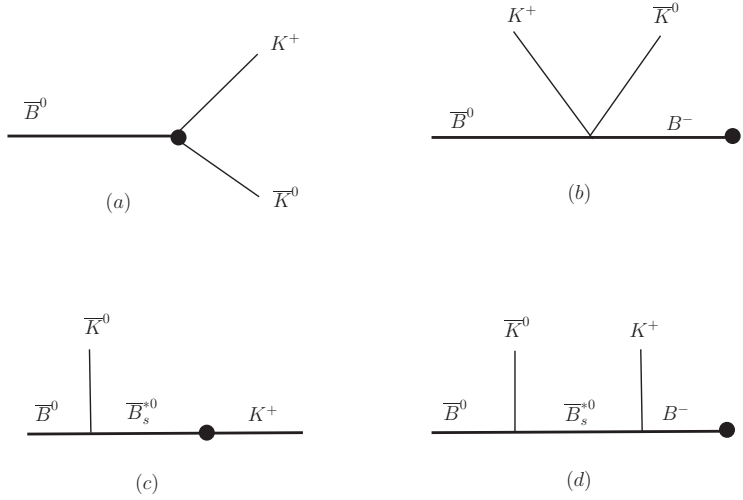,width=10cm}
\vspace{00.5cm}
    \caption[]{\small Point-like and pole diagrams responsible for the $\ov B^0\to K^+\ov K^0$
    matrix element induced by the current $\bar u\gamma_\mu(1-\gamma_5)b$,
    where the symbol $\bullet$ denotes an insertion of the current.}
\end{figure}

A direct calculation indicates that the branching ratio of $\ov
B^0\to K^+K^-\ov K^0$ arising from the current-induced process
alone is already at the level of $77\times 10^{-6}$ which exceeds
the measured total branching  ratio of $25\times 10^{-6}$ (see
Table \ref{tab:BRexpt}). The issue has to do with the
applicability of HMChPT. In order to apply this approach, two of
the final-state pseudoscalars ($K^+$ and $\ov K^0$ in this
example) have to be soft. The momentum of the soft pseudoscalar
should be smaller than the chiral symmetry breaking scale
$\Lambda_\chi$ of order $0.83-1.0$ GeV. For 3-body charmless $B$
decays, the available phase space where chiral perturbation theory
is applicable is only a small fraction of the whole Dalitz plot.
Therefore, it is not justified to apply chiral and heavy quark
symmetries to a certain kinematic region and then generalize it to
the region beyond its validity. If the soft meson result is
assumed to be the same in the whole Dalitz plot, the decay rate
will be greatly overestimated.

In \cite{CCSKKK,Cheng:2002qu} we have tried to circumvent the
aforementioned problem by applying HMChPT only to the strong
vertex and use the form factors to describe the weak vertex.
Moreover, we introduced a form factor to take care of the
off-shell effect. For example, Fig. 1(c) can be evaluated by
considering the strong interaction $\ov B {}^0\to \ov K {}^0 \ov
B_s^*$ followed by the weak transition $\ov B_s^*\to K^+$ and the
result is \cite{Cheng:2002qu}
 \be \label{eq:BKKpole}
 A_{Fig. 1(c)}
 &=&\frac{f_K}{f_\pi}\frac{g\sqrt{m_B
 m_{B_s^*}}}{s_{23}-m^2_{B_s^*}}
 F(s_{23},m_{B_s^*}) F_1^{B_sK}(m_3^2)
 \bigg[m_B+\frac{s_{23}}{m_B}-m_B\frac{m_B^2-s_{23}}{m_3^2}
 \bigg(1-\frac{F_0^{B_sK}(m_3^2)}{F_1^{B_sK}(m_3^2)}\bigg)\bigg]\non\\
 &&\qquad\times\bigg[m_1^2+m_3^2-s_{13}+\frac{(s_{23}-m_2^2+m_3^2)(m_B^2-s_{23}-m_1^2)}{2
 m_{B_s^*}^2}\bigg],
 \en
where $F^{B_s K}_{0,1}$ are the $B_s\to K$ weak transition from
factors in the standard convention \cite{BSW} and we have
introduced a form factor $F(s_{23},m_{B_s^*})$ to take into
account the off-shell effect of the $B_s^*$ pole \cite{CCSKKK}. It
is parameterized as
$F(s_{23},m_{B_s^*})=(\Lambda^2-m_{B_s^*}^2)/(\Lambda^2-s_{23})$
with the cut-off parameter $\Lambda$ chosen to be
$\Lambda=m_{B^*_s}+\Lambda_{\rm QCD}$. Needless to say, this
parametrization of the form factor is somewhat arbitrary.
Moreover, the nonresonant contribution thus calculated is too
small compared to experiment.

The Dalitz plot analysis of $\ov B^0\to K^+K^-\ov K^0$ has been
recently performed by BaBar \cite{BaBarKpKmK0}. In the BaBar
analysis, a phenomenological parametrization of the non-resonant
amplitudes is described by
 \be \label{eq:ANR}
A_{\rm NR}=(c_{12}e^{i\phi_{12}}e^{-\alpha
s_{12}^2}+c_{13}e^{i\phi_{13}}e^{-\alpha
s_{13}^2}+c_{23}e^{i\phi_{23}}e^{-\alpha s_{23}^2})(1+b_{\rm
NR}e^{i(\beta+\delta_{\rm NR})}),
 \en
and resonant terms are described by
 \be
A_{\rm R}=\sum_r c_r(1+b_r)f_r e^{i(\phi_r+\delta_r+\beta)},
\qquad \bar {A}_{\rm R}=\sum_r c_r(1-b_r)f_r
e^{i(\phi_r-\delta_r+\beta)}.
 \en
The BaBar results for isobar amplitudes, phases and fractions from
the fit to the $B^0\to K^+K^-K^0$ are summarized in Table
\ref{tab:ExpKpKmK0}. It is evident that this decay is dominated by
the nonresonant background. For our purpose, we will parametrize
the current-induced nonresonant amplitude Eq. (\ref{eq:AHMChPT})
as
 \be \label{eq:ADalitz}
  A_{\rm current-ind}=A_{\rm current-ind}^{\rm
  HMChPT}\,e^{-\alpha_{_{\rm NR}}
p_B\cdot(p_1+p_2)}e^{i\phi_{12}},
 \en
so that the HMChPT results are recovered in the chiral limit
$p_1,~p_2\to 0$. That is, the nonresonant amplitude in the soft
meson region is described by HMChPT, but its energy dependence
beyond the chiral limit is governed by the exponential term
$e^{-\alpha_{_{\rm NR}} p_B\cdot(p_1+p_2)}$.  In what follows, we
shall use the tree-dominated $B^-\to\pi^+\pi^-\pi^-$ decay data to
fix $\alpha_{_{\rm NR}}$, which turns out to be
 \be \label{eq:alpha}
 \alpha_{_{\rm NR}}=0.103^{+0.018}_{-0.011}\,{\rm GeV}^{-2}.
 \en
This is very close to the naive expectation of $\alpha_{_{\rm
NR}}\sim {\cal O}(1/(2m_B\Lambda_\chi))$ based on the dimensional
argument. The phase $\phi_{12}$ of the nonresonant amplitude in
the $(K^+\ov K^0)$ system will be set to zero for simplicity.

\begin{table}[h]
\caption{\small BaBar results for isobar amplitudes, phases, and
fractions from the fit to the $B^0\to K^+K^-K^0$
\cite{BaBarKpKmK0}. Three rows for non-resonant contribution
correspond to coefficients of exponential functions in
Eq.~(\ref{eq:ANR}), while the fraction is given for the combined
amplitude. For the nonresonant decay mode in $K^+K^-$, the
amplitude $c_{12}$ and the phase $\phi_{12}$ in Eq. (\ref{eq:ANR})
are fixed to be one and zero, respectively. Errors are statistical
only. } \label{tab:ExpKpKmK0} \center
\begin{tabular}{|lr|rrr|}
\hline \hline
\multicolumn{2}{|l|}{~~Decay }            & ~~~~Amplitude $c_r$       & ~~~~~~~Phase $\phi_r$   & ~~~~Fraction (\%)~~\\
\hline
\multicolumn{2}{|l|}{$\phi(1020)K^0$}   & $  0.0085\pm 0.0010$       &       $ -0.016 \pm 0.234$   &       $12.5 \pm 1.3$         \\
\multicolumn{2}{|l|}{$f_0(980)K^0$} & $  0.622 \pm 0.046$         &       $ -0.14 \pm 0.14$       &       $40.2 \pm 9.6$    \\
\multicolumn{2}{|l|}{$X_0(1550)K^0$}    & $  0.114 \pm 0.018$       &       $ -0.47 \pm 0.20$   &       $4.1 \pm 1.3$      \\
&$(K^+K^-)_{\rm NR}K^0$                & 1 (fixed)                 &       0 (fixed)        &                              \\
&$(K^+K^0)_{\rm NR}K^-$                 & $  0.33 \pm 0.07$         &       $  1.95  \pm 0.27$     &       $112.0 \pm 14.9$     \\
&$(K^-K^0)_{\rm NR}K^+$                 & $  0.31 \pm 0.08$     &       $ -1.34  \pm 0.37$    &                           \\
\hline
\multicolumn{2}{|l|}{$\chi_{c0}(1P)K^0$}    & $  0.0306\pm 0.00649$      &       $ ^{~~0.81}_{-2.33} \pm 0.54$    &       $3.0 \pm 1.2$       \\
\multicolumn{2}{|l|}{$D^+K^-$}          & $  1.11\pm 0.17$      &               --          &       $3.6 \pm 1.5$          \\
\multicolumn{2}{|l|}{$D_s^+K^-$}          & $  0.76\pm 0.14$      &               --          &       $1.8 \pm 0.6$        \\
\hline \hline
\end{tabular}
\end{table}

For the transition amplitude, we need to evaluate the 2-kaon
creation matrix element which can be expressed in terms of
time-like kaon current form factors as
 \be \label{eq:KKweakff}
 \la K^+(p_{K^+}) K^-(p_{K^-})|\bar q\gamma_\mu q|0\ra
 &=& (p_{K^+}-p_{K^-})_\mu F^{K^+K^-}_q,
 \non\\
 \la K^0(p_{K^0}) \ov K^0(p_{\bar K^0})|\bar q\gamma_\mu q|0\ra
 &=& (p_{K^0}-p_{\bar K^0})_\mu F^{K^0\bar K^0}_q.
 \en
The weak vector form factors $F^{K^+K^-}_q$ and $F^{K^0\bar
K^0}_q$ can be related to the kaon electromagnetic (e.m.) form
factors $F^{K^+K^-}_{em}$ and $F^{K^0\bar K^0}_{em}$ for the
charged and neutral kaons, respectively. Phenomenologically, the
e.m. form factors receive resonant and nonresonant contributions
and can be expressed by
 \be \label{eq:KKemff}
 F^{K^+K^-}_{em}= F_\rho+F_\omega+F_\phi+F_{NR}, \qquad
 F^{K^0\bar K^0}_{em}= -F_\rho+F_\omega+F_\phi+F_{NR}'.
 \en
It follows from Eqs. (\ref{eq:KKweakff}) and (\ref{eq:KKemff})
that
 \be
 F^{K^+K^-}_u&=&F^{K^0\bar K^0}_d=F_\rho+3 F_\omega+\frac{1}{3}(3F_{NR}-F'_{NR}),
 \non\\
 F^{K^+K^-}_d&=&F^{K^0\bar K^0}_u=-F_\rho+3 F_\omega,
 \non\\
 F^{K^+K^-}_s&=&F^{K^0\bar K^0}_s=-3 F_\phi-\frac{1}{3}(3 F_{NR}+2F'_{NR}),
 \label{eq:FKKisospin}
 \en
where use of isospin symmetry has been made.

The resonant and nonresonant terms in Eq. (\ref{eq:KKemff}) can be
parametrized as
 \be
 F_{h}(s_{23})=\frac{c_h}{m^2_h-s_{23}-i m_h \Gamma_h},
 \qquad
 F^{(\prime)}_{NR}(s_{23})=\left(\frac{x^{(\prime)}_1}{s_{23}}
 +\frac{x^{(\prime)}_2}{s_{23}^2}\right)
 \left[\ln\left(\frac{s_{23}}{\tilde\Lambda^2}\right)\right]^{-1},
 \en
with $\tilde\Lambda\approx 0.3$ GeV. The expression for the
nonresonant form factor is motivated by the asymptotic constraint
from pQCD, namely, $F(t)\to (1/t)[\ln(t/\tilde \Lambda^2)]^{-1}$
in the large $t$ limit \cite{Brodsky}. The unknown parameters
$c_h$, $x_i$ and $x'_i$ are fitted from the kaon e.m. data, giving
the best fit values (in units of GeV$^2$ for $c_h$) ~\cite{DKK}:
\begin{equation}
\begin{array}{lll}
c_\rho=3c_\omega=c_\phi=0.363,
  & c_{\rho(1450)}=7.98\times 10^{-3},\ \
  & c_{\rho(1700)}=1.71\times10^{-3},\ \
\\
c_{\omega(1420)}=-7.64\times 10^{-2},
  & c_{\omega(1650)}=-0.116,
  & c_{\phi(1680)}=-2.0\times10^{-2},
\\
\end{array}
\label{eq:cj}
\end{equation}
and
\begin{eqnarray}
x_1=-3.26~{\rm GeV}^2, \qquad x_2=5.02~{\rm GeV}^4,
 \qquad x'_1=0.47~{\rm GeV}^2,
 \qquad x'_2=0.
\label{eq:xy}
\end{eqnarray}
Note that the form factors $F_{\rho,\omega,\phi}$ in
Eqs.~(\ref{eq:KKemff}) and (\ref{eq:FKKisospin}) include the
contributions from the vector mesons
$\rho(770),\,\rho(1450),\,\rho(1700)$,
$\omega(782),\,\omega(1420),\,\omega(1650),$ $\phi(1020)$ and
$\phi(1680)$.
It is interesting to note that (i) the fitted values of $c_{V}$
are very close to the vector meson dominance expression
$g_{_{V\gamma}} g_{VKK}$ for $V=\rho,\omega,\phi$~\cite{DM2,PDG},
where $g_{_{V\gamma}}$ is the e.m. coupling of the vector meson
defined by $\la V|j_{em}|0\ra=g_{V\gamma}\vp^*_V$ and $g_{VKK}$ is
the $V\to KK$ strong coupling with $-g_{\phi K^+K^-}\simeq g_{\rho
K^+K^-}/\sqrt2= g_{\omega K^+K^-}/\sqrt2\simeq3.03$,
and (ii) the vector-meson pole contributions alone yield
$F^{K^+K^-}_{u,s}(0)\approx 1,-1$ and $F^{K^+K^-}_d(0)\approx 0$
as the charged kaon does not contain the valence $d$ quark. The
matrix element for the current-induced decay process then has the
expression
 \be
 \la \ov K {}^0(p_1)|(\bar s b)_{V-A}|\ov B {}^0\ra \la K^+(p_2)
 K^-(p_3)|(\bar q q)_{V-A}|0\ra
 =(s_{12}-s_{13}) F_1^{BK}(s_{23}) F^{K^+K^-}_q (s_{23}).
  \en

We also need to specify the 2-body matrix elements $\la K^+
K^-|\bar s s|0\ra\la\ov K^0|\bar sb|\ov B^0\ra$ induced from the
scalar densities. The use of the equation of motion leads to
 \be
 \la\ov K^0(p_1)|\bar sb|\ov B^0(p_B)\ra=\,{m_B^2-m_K^2\over
 m_b-m_s}F_0^{BK}(s_{23}).
 \en
The matrix element $\la K^+ K^-|\bar s s|0\ra$ receives resonant
and non-resonant contributions:
 \be \label{eq:KKssme}
 \la K^+(p_2) K^-(p_3)|\bar s s|0\ra
 &\equiv& f^{K^+K^-}_s(s_{23})=\sum_{i}\frac{m_{{f_0}_i} \bar f^s_{{f_0}_i} g^{{f_0}_i\to K^+K^-}}{m_{{f_0}_i}^2-s_{23}-i
 m_{{f_0}_i}\Gamma_{{f_0}_i}}+f_s^{NR},
 \non\\
 f_s^{NR}&=&\frac{v}{3}(3 F_{NR}+2F'_{NR})+\sigma_{_{\rm NR}}
 e^{-\alpha\,s_{23}},
 \en
where ${f_0}_i$ denote the generic $f_0$-type scalar mesons,
${f_0}_i=f_0(980),f_0(1370),f_0(1500),X_0(1550),\cdots$, the
scalar decay constant $\bar f_{{f_0}_i}^s$ is defined by $\la
{f_0}_i|\bar s s|0\ra=m_{{f_0}_i} \bar f^s_{{f_0}_i}$ [see Eq.
(\ref{eq:decayc})], $g^{{f_0}_i\to K^+K^-}$ is the ${f_0}_i\to
K^+K^-$ strong coupling, and the nonresonant terms are related to
those in $F_s^{K^+K^-}$ through the equation of motion. The
presence of the nonresonant $\sigma_{_{\rm NR}}$ term will be
explained shortly. The main scalar meson pole contributions are
those that have dominant $s\bar s$ content and large coupling to
$K\ov K$. We consider the scalar mesons $f_0(980)$ and $X_0(1550)$
(denoted as $f_X(1500)$ by Belle) which are supposed to have the
largest couplings with the $K\ov K$ pair. Note that the nature of
the broad state $X_0(1550)$ observed by BaBar and Belle, for
example, what is its relation with $f_0(1500)$, is not clear. To
proceed with the numerical calculations, we shall use
$g^{f_0(980)\to K^+K^-}=4.3$~GeV,\footnote{This is different from
the coupling $g^{f_0(980)\to K^+K^-}=1.5$ GeV originally employed
in \cite{CCSKKK}. The coupling $g^{f_0(980)\to \pi^+\pi^-}\sim
1.33$ GeV can be fixed from a recent Belle measurement of
$\Gamma(f_0(980)\to\pi^+\pi^-)$ [see Eq. (\ref{eq:g})]. Using the
BES result $(g^{f_0(980)\to KK}/g^{f_0(980)\to
\pi\pi})^2=4.21\pm0.25\pm0.21$ \cite{BES}, one can deduce that
$g^{f_0(980)\to KK}=2.7\pm0.6$~GeV. In this work, we found that a
slightly large coupling $g^{f_0(980)\to KK}$ will give better
numerical results.}
$g^{X_0(1550)\to K^+K^-}=1.4$~GeV, $\Gamma_{f_0(980)}=80$~MeV,
$\Gamma_{X_0(1550)}=0.257$~GeV~\cite{BaBarKpKpKm}, $\bar
f_{f_0(980)}(\mu=m_b/2)\simeq 0.46$~GeV~\cite{Cheng:2005ye} and
$\bar f_{f_0(1530)}\simeq 0.30$ GeV.
The sign of the resonant terms is fixed by $f_s^{K^+ K^-}(0)=v$
from a chiral perturbation theory calculation (see, for example,
\cite{Cheng:1988va}).
It should be stressed that although the nonresonant contributions
to $f_s^{KK}$ and $F_s^{KK}$ are related through the equation of
motion, the resonant ones are different and  not related {\it a
priori}. As stressed in \cite{CCSKKK}, to apply the equation of
motion, the form factors should be away from the resonant region.
In the presence of the resonances, we thus need to introduce a
nonresonant term characterized by the parameter $\sigma_{\rm NR}$
in Eq. (\ref{eq:KKssme}) which will be specified later. The
parameter $\alpha$ appearing in the same equation should be close
to the value of $\alpha_{_{\rm NR}}$ given in Eq.
(\ref{eq:alpha}). We will use the experimental measurement
$\alpha=(0.14\pm0.02)\,{\rm GeV}^{-2}$ \cite{BaBarKpKmK0}.

As noticed before, the matrix elements $\la K^+K^-|(\bar
db)_\vma|\ov B^0\ra$ and $\la K^+K^-|\bar d(1-\gamma_5)b|\ov
B^0\ra$ are included in Eq. (\ref{eq:AKpKmK0}) as they receive
intermediate scalar pole contributions. More explicitly,
 \be
 \la K^+(p_2)K^-(p_3)|(\bar db)_\vma|\ov B^0\ra^R =
\sum_{i}\frac{g^{{f_0}_i\to K^+K^-}}{m_{{f_0}_i}^2-s_{23}-i
 m_{{f_0}_i} \Gamma_{{f_0}_i}}\la {f_0}_i|(\bar
db)_\vma|\ov B^0\ra.
 \en
Hence,
 \be
&& \la \ov K^0(p_1)|(\bar sd)_\vma|0\ra\la K^+(p_2)K^-(p_3)|(\bar
db)_\vma|\ov B^0\ra^R \non \\ &=& \sum_{i}\frac{g^{{f_0}_i\to
K^+K^-}}{m_{{f_0}_i}^2-s_{23}-i
 m_{{f_0}_i}\Gamma_{{f_0}_i}} f_KF_0^{B{f_0}_i^d}(m_K^2)(m_B^2-m_{{f_0}_i}^2).
 \en
The superscript $u$ of the form factor $F_0^{B{f_0}_i^u}$ reminds
us that it is the $u\bar u$ quark content that gets involved in
the $B$ to ${f_0}_i$ form factor transition. In short, the
relevant $f_0(980)$ pole contributions to $\ov B^0\to K^+K^-\ov
K^0$ are
 \be
 \la \ov K^0K^+K^-|T_p|\ov
B^0\ra_{f_0} &=& {g^{f_0(980)\to K^+K^-}\over
m_{f_0}^2-s_{23}-im_{f_0}\Gamma_{f_0}}\Bigg\{ 2{m_{f_0}\over
m_b}\bar
f_{f_0}^sF_0^{BK}(m_{f_0}^2)(m_B^2-m_K^2)\left(a_6^p-{1\over
2}a_8^p\right) \non \\
&+& f_KF_0^{Bf_0^d}(m_K^2)(m_B^2-m_{f_0}^2)\left[a_4^p-{1\over
2}a_{10}^p-(a_6^p-{1\over 2}a_8^p)r_\chi^K\right] \Bigg\},
 \en
where we have employed Eq. (\ref{eq:KKssme}) and applied equations
of motion to the matrix elements $\la \ov K^0|\bar s\gamma_5
d|0\ra\la K^+K^-|\bar d\gamma_5b|\ov B^0\ra$. Comparing this
equation with Eq. (A6) of \cite{CCY}, we see that the expression
inside $\{\cdots\}$ is identical to that of $\ov B^0\to
f_0(980)\ov K^0$, as it should be.

We digress for a moment to discuss the wave function of the
$f_0(980)$. What is the quark structure of the light scalar mesons
below or near 1 GeV has been quite controversial. In this work we
shall consider the conventional $q\bar q$ assignment for the
$f_0(980)$. In the naive quark model, the flavor wave functions of
the $f_0(980)$ and $\sigma(600)$ read
 \be
 \sigma={1\over \sqrt{2}}(u\bar u+d\bar d), \qquad\qquad f_0= s\bar
 s,
 \en
where the ideal mixing for $f_0$ and $\sigma$ has been assumed. In
this picture, $f_0(980)$ is purely an $s\bar s$ state. However,
there also exist some experimental evidences indicating that
$f_0(980)$ is not purely an $s\bar s$ state. First, the
observation of $\Gamma(J/\psi\to f_0\omega)\approx {1\over
2}\Gamma(J/\psi\to f_0\phi)$ \cite{PDG} clearly indicates the
existence of the non-strange and strange quark content in
$f_0(980)$. Second, the fact that $f_0(980)$ and $a_0(980)$ have
similar widths and that the $f_0$ width is dominated by $\pi\pi$
also suggests the composition of $u\bar u$ and $d\bar d$ pairs in
$f_0(980)$; that is, $f_0(980)\to\pi\pi$ should not be OZI
suppressed relative to $a_0(980)\to\pi\eta$. Therefore, isoscalars
$\sigma(600)$ and $f_0$ must have a mixing
 \be
 |f_0(980)\ra = |s\bar s\ra\cos\theta+|n\bar n\ra\sin\theta,
 \qquad |\sigma(600)\ra = -|s\bar s\ra\sin\theta+|n\bar n\ra\cos\theta,
 \en
with $n\bar n\equiv (\bar uu+\bar dd)/\sqrt{2}$. Experimental
implications for the $f_0\!-\!\sigma$ mixing angle have been
discussed in detail in \cite{ChengDSP}. It is found that $\theta$
lies in the ranges of $25^\circ<\theta<40^\circ$ and
$-40^\circ<\theta< -15^\circ$ (or $140^\circ<\theta< 165^\circ$).
Note that the phenomenological analysis of the radiative decays
$\phi\to f_0(980)\gamma$ and $f_0(980)\to\gamma\gamma$ favors a
solution of the $\theta$ to be negative (or in the second
quadrant). In this work, we shall use $\theta=-25^\circ$.


Finally, the matrix elements involving 3-kaon creation are given
by~\cite{Cheng:2002qu}
 \be \label{eq:KKKme}
 &&\hspace{-0.5cm}\la \ov K {}^0(p_1) K^+(p_2) K^-(p_3)|(\bar s d)_{V-A}|0\ra\la
 0|(\bar d b)_{V-A}|\ov B {}^0\ra
 \approx  0, \\
 &&\hspace{-0.5cm}\la \ov K {}^0(p_1) K^+(p_2) K^-(p_3)|\bar s\gamma_5
 d|0\ra\la
 0|\bar d\gamma_5 b|\ov B {}^0\ra=v\frac{ f_B m_B^2}{f_\pi m_b}
 \left(1-\frac{s_{13}-m_1^2-m_3^2}{m_B^2-m_K^2}\right)F^{KKK}(m_B^2),
 \non
 \en
where
 \be \label{eq:v}
 v=\frac{m_{K^+}^2}{m_u+m_s}=\frac{m_K^2-m_\pi^2}{m_s-m_d},
 \en
characterizes the quark-order parameter $\la \bar q q\ra$ which
spontaneously breaks the chiral symmetry. Both relations in Eq.
(\ref{eq:KKKme}) are originally derived in the chiral limit
\cite{Cheng:2002qu} and hence the quark masses appearing in Eq.
(\ref{eq:v}) are referred to the scale $\sim$ 1 GeV . The first
relation reflects helicity suppression which is expected to be
even more effective for energetic kaons. For the second relation,
we introduce the form factor $F^{KKK}$ to extrapolate the chiral
result to the physical region. Following \cite{Cheng:2002qu} we
shall take $F^{KKK}(q^2)=1/[1-(q^2/\Lambda^2_\chi)]$ with
$\Lambda_\chi=0.83$~GeV being a chiral symmetry breaking scale.

\begin{figure}[t]
  \centerline{\epsfig{figure=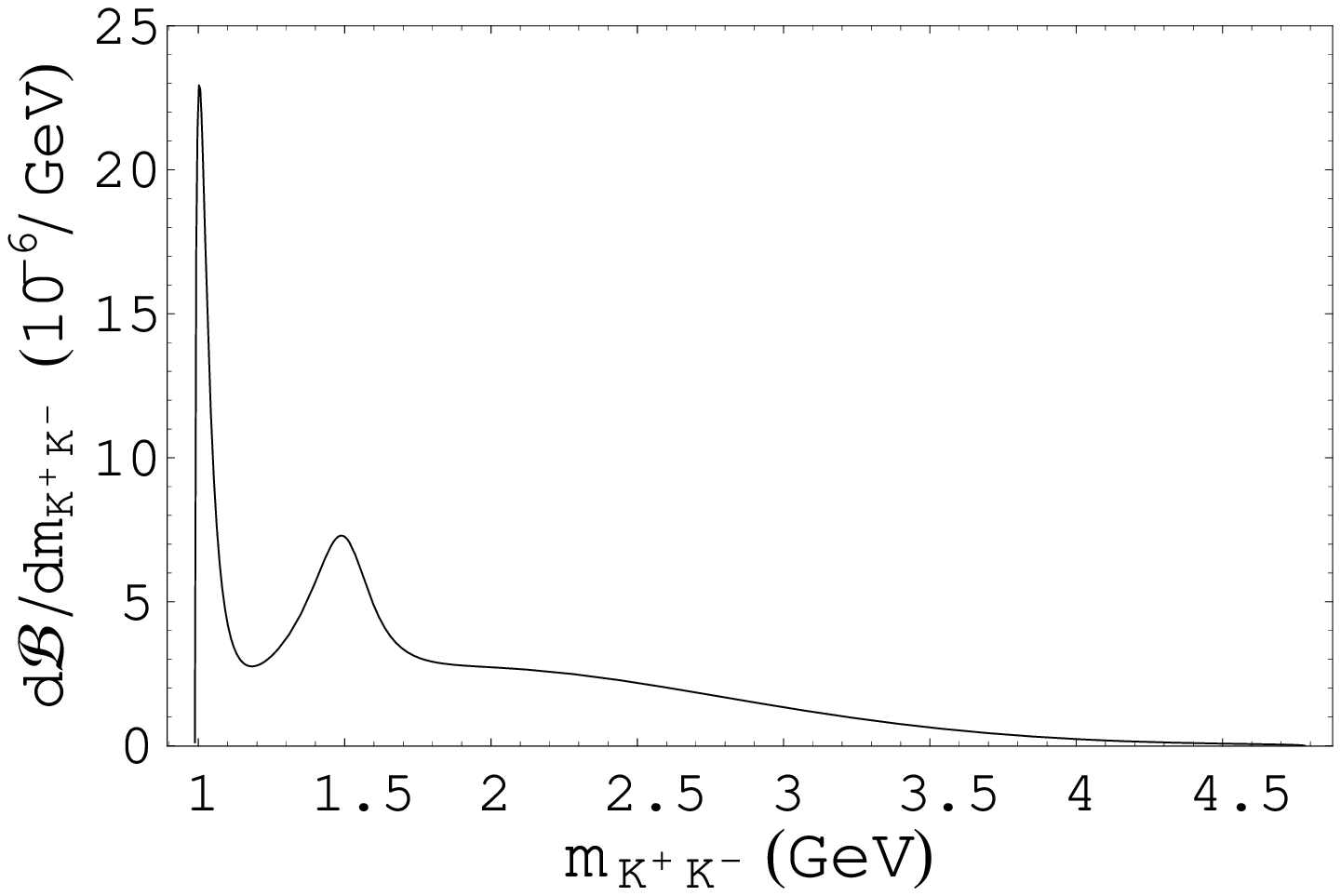,width=5.0cm}
              \hspace{0.2cm}
              \epsfig{figure=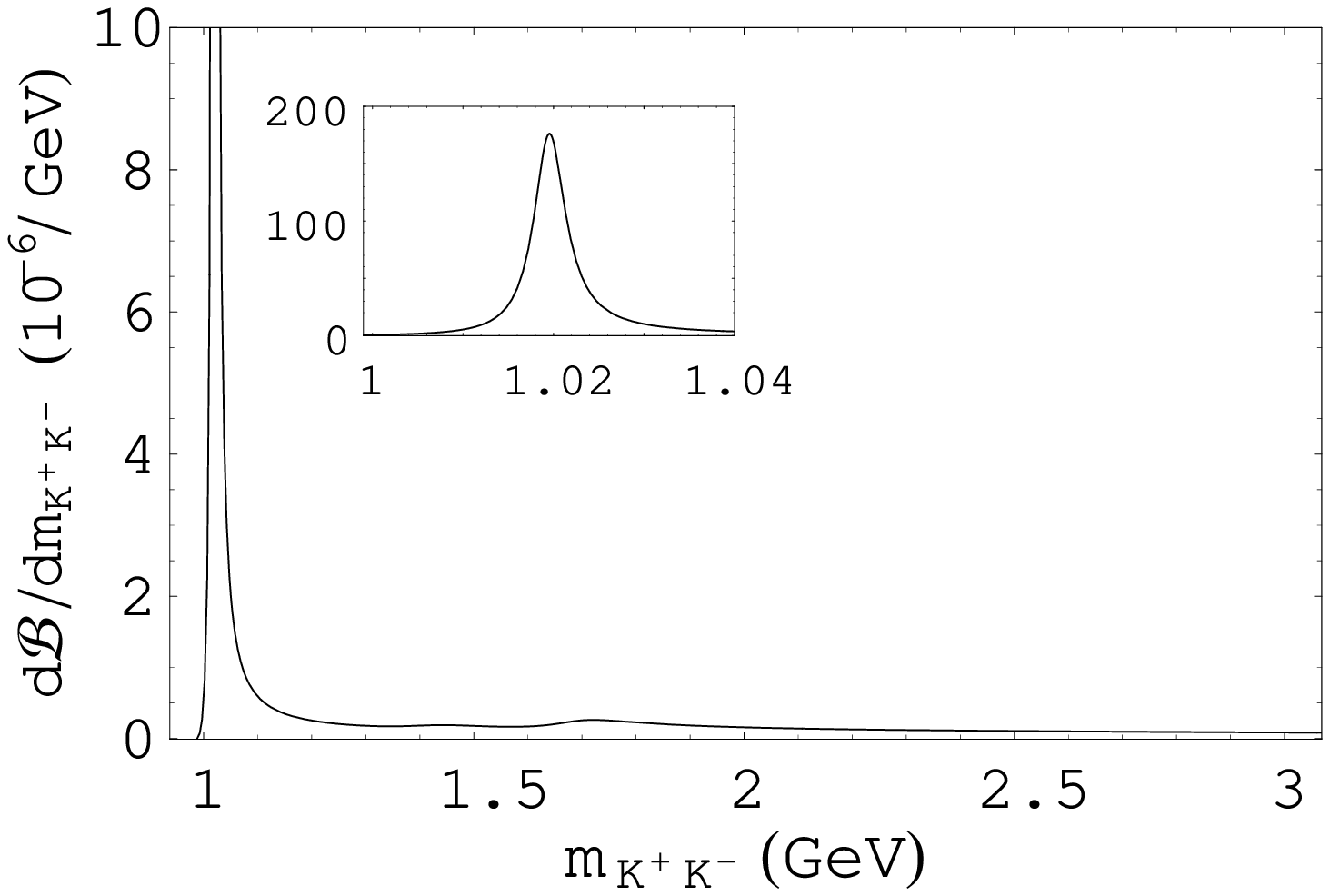,width=5.2cm}
              \hspace{0.2cm}
              \epsfig{figure=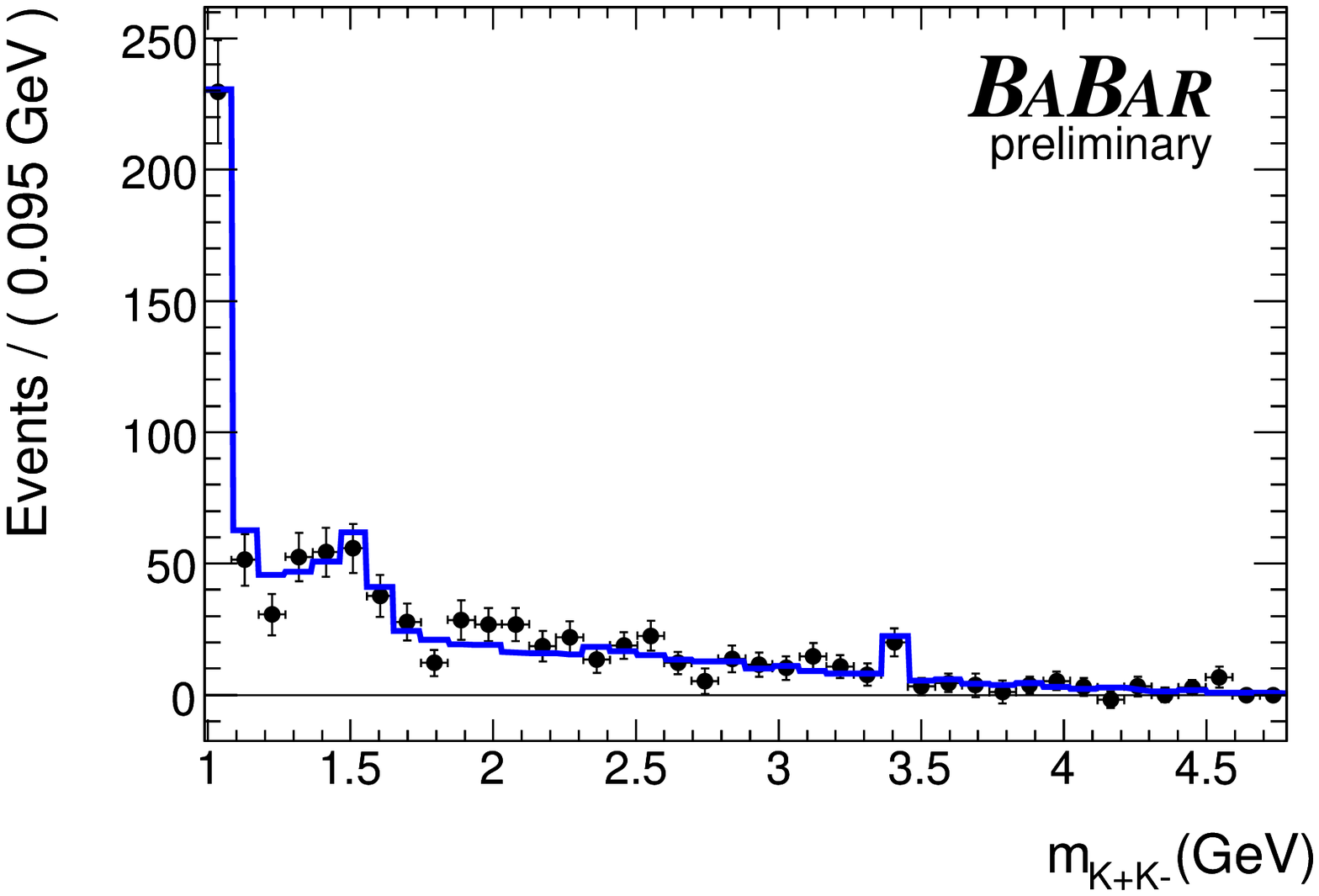,width=5.0cm}
              }
  \centerline{(a)
              \hspace{5.3cm}
              (b)
              \hspace{5.5cm}
              (c)
              }
    \caption{\small The $K^+ K^-$ mass spectra for
    $\overline B {}^0\to K^+ K^- K_S$ decay from (a) $CP$-even and (b) $CP$-odd contributions.
    The insert in (b) is for the $\phi$ region. The
full $K^+K^-K_S$ spectrum, which is the sum of $CP$-even and
$CP$-odd parts, measured by BaBar \cite{BaBarKpKmK0} is depicted
in (c).}\label{fig:spectrum}
\end{figure}

To proceed with the numerical calculations, we need to specify the
input parameters. The relevant CKM matrix elements, decay
constants, form factors, the effective Wilson coefficients $a_i^p$
and the running quark masses are collected in Appendix B. As for
the parameter $\sigma_{_{\rm NR}}$ in Eq. (\ref{eq:KKssme}), in
principle we can set its phase $\phi_\sigma$ to zero and use the
measured $K_SK_SK_S$ rate, namely, $\B(\ov B^0\to
K_SK_SK_S)=(6.2\pm0.9)\times 10^{-6}$ \cite{HFAG}, to fix the
parameter $\sigma_{_{\rm NR}}$ and then use the data obtained from
the Dalitz plot analysis to determine the strong phases $\phi_r$
for resonant amplitudes. However, in doing so one needs the data
of invariant mass spectra. In the absence of such information,
instead we will treat $\phi_\sigma$ as a free parameter and do not
assign any other strong phases to the resonant amplitudes except
for those arising from the Breit-Wigner formalism. It turns out
that if $\phi_\sigma$ is small, the $K^+K^-$ mass spectrum in $\ov
B^0\to K^+K^-K_S$ will have a prominent hump at the invariant mass
$m_{K^+K^-}=3$ GeV, which is not seen experimentally (see Fig.
\ref{fig:spectrum}(c)). We found that $\phi_\sigma\approx \pi/4$
will yield $K^+K^-$ mass spectrum consistent with the data
 \be \label{eq:sigma}
  \sigma_{_{\rm NR}}= e^{i\pi/4}\left(3.36^{+1.12}_{-0.96}\right)\,{\rm GeV}.
 \en
Note that the phase of $\sigma_{_{\rm NR}}$ is consistent with the
BaBar measurement shown in Table \ref{tab:ExpKpKmK0}, namely,
$\phi_\sigma^{\rm BaBar}=1.19\pm0.37$.

The calculated branching ratios of resonant and nonresonant
contributions to $\ov B^0\to K^+K^-\ov K^0$ are summarized in
Table \ref{tab:KpKmK0}. The theoretical errors shown there are
from the uncertainties in (i) the parameter $\alpha_{\rm NR}$
which governs the momentum dependence of the nonresonant
amplitude, (ii) the strange quark mass $m_s$, the form factor
$F^{BK}_0$ and the nonresonant parameter $\sigma_{_{\rm NR}}$, and
(iii) the unitarity angle $\gamma$.

In QCD calculations based on a heavy quark expansion, one faces
uncertainties arising from power corrections such as annihilation
and hard-scattering contributions. For example, in QCD
factorization, there are large theoretical uncertainties related to
the modelling of power corrections corresponding to weak
annihilation effects and the chirally-enhanced power corrections to
hard spectator scattering. Even for two-body $B$ decays, power
corrections are of order (10-20)\% for tree-dominated modes, but
they are usually bigger than the central values for
penguin-dominated decays. Needless to say, $1/m_b$ power
corrections for three-body decays may well be larger. However, as
stressed in Introduction, in this exploratory work we use the
phenomenological factorization model rather than in the established
theories based on a heavy quark expansion. Consequently,
uncertainties due to power corrections, at this stage, are not
included in our calculations, by assumption. In view of such
shortcomings we must emphasize that the additional errors due to
such model dependent assumptions may be sizable.

From Table \ref{tab:KpKmK0} we see that the predicted rates for
resonant and nonresonant components are consistent with experiment
within errors. The nonresonant contribution arises dominantly from
the transition process (88\%) via the scalar-density-induced
vacuum to $K\bar K$ transition, namely, $\la K^+K^-|\bar ss|0\ra$,
and slightly from the current-induced process (3\%). Therefore, it
is natural to conjecture that nonresonant decays could also play a
prominent role in other penguin dominated 3-body $B$ decays.

\begin{table}[t]
\caption{Branching ratios (in units of $10^{-6}$) of resonant and
nonresonant (NR) contributions to $\ov B^0\to K^+K^-\ov K^0$.
Theoretical errors correspond to the uncertainties in (i)
$\alpha_{_{\rm NR}}$, (ii) $m_s$, $F^{BK}_0$ and $\sigma_{_{\rm
NR}}$, and (iii) $\gamma=(59\pm7)^\circ$. We do not have $1/m_b$
power corrections within this model. However, systematic errors due
to model dependent assumptions may be sizable and are not included
in the error estimates that we give. Experimental results are taken
from Table \ref{tab:ExpKpKmK0}.   }
\begin{ruledtabular} \label{tab:KpKmK0}
\begin{tabular}{l l  l}
 Decay mode~~ & BaBar \cite{BaBarKpKmK0}  & Theory  \\ \hline
 $\phi \ov K^0$ & $2.98\pm0.45$ & $2.6^{+0.0+0.5+0.0}_{-0.0-0.4-0.0}$ \\
 $f_0(980)\ov K^0$ & $9.57\pm2.51$ & $5.8^{+0.0+0.1+0.0}_{-0.0-0.5-0.0}$ \\
 $X_0(1550)K^-$ & $0.98\pm0.33$ & $0.93^{+0.00+0.16+0.00}_{-0.00-0.15-0.00}$ \\
 NR & $26.7\pm4.6$ & $18.1^{+0.6+5.1+0.2}_{-0.7-3.8-0.2}$ \\
\hline
 total & $23.8\pm2.0\pm1.6$  &  $19.8^{+0.4+0.5+0.1}_{-0.4-0.4-0.2}$ \\
\end{tabular}
\end{ruledtabular}
\end{table}

The $K^+K^-K_S$ mode is an admixture of $CP$-even and $CP$-odd
components. By excluding the major $CP$-odd contribution from
$\phi K_S$, the 3-body $K^+K^-K_S$ final state is primarily
$CP$-even. The $K^+K^-$ mass spectra of the $\overline B {}^0\to
K^+ K^- K_S$ decay from $CP$-even and $CP$-odd contributions are
shown in Fig.~\ref{fig:spectrum}.  For the $CP$-even spectrum,
there are peaks at the threshold and $m_{K^+K^-}=1.5$ GeV region.
The threshold enhancement arises from the $f_0(980) K_S$ and the
nonresonant $f_S^{K^+K^-}$ contributions [see Eq.
(\ref{eq:KKssme})]. \footnote{In our previous work \cite{CCSKKK}
we have argued that the spectrum should have a peak at the large
$m_{K^+ K^-}$ end. This is because we have introduced an
additional nonresonant contribution to the $\omega_-$ parameter
parametrized as $\omega^{NR}_-=\kappa\, \frac{2p_B\cdot
p_2}{s^2_{12}}$ and employed the $B^-\to D^0 K^0 K^-$ data and
applied isospin symmetry to the $\ov B\to K\ov K$ matrix elements
to determine the unknown parameter $\kappa$. Since this
nonresonant term favors a small $m_{K^+ K_S}$ region, a peak of
the spectrum at large $m_{K^+K^-}$ is thus expected. However, such
a bump is not seen experimentally \cite{BaBarKpKmK0}. In this work
we will no longer consider this term.}
For the $CP$-odd spectrum, the peak on the lower end corresponds
to the $\phi K_S$ contribution, which is also shown in the insert.
The $b\to u$ transition is governed by the current-induced process
$\la \ov B^0\to K^+\ov K^0\ra\times \la 0\to K^-\ra$ [see Eq.
(\ref{eq:AKpKmK0})]. From Eq. (\ref{eq:ADalitz}) it is clear that
the $b\to u$ amplitude prefers a small invariant mass of $K^+$ and
$\ov K^0$ and hence a large invariant mass of $K^+$ and $K^-$. In
contrast, the $b\to c$ amplitude prefers a small $s_{23}$.
Consequently, their interference is largely suppressed. The full
$K^+K^-K_S$ spectrum, which is the sum of the $CP$-even and the
$CP$-odd parts, has been measured by BaBar
[Fig.~\ref{fig:spectrum}(c)]. It clearly shows the phenomenon of
threshold enhancement and the scalar resonances $X_0(1550)$ and
$\chi_{c0}$.

The decay $\ov B^0\to K_SK_SK_S$ is a pure penguin-induced mode
[cf. Eq. (\ref{eq:AK0K0K0})] and it receives intermediate pole
contributions only from the iso-singlet scalar mesons such as
$f_0(980)$. Just like other $KKK$ modes, this decay is governed by
the nonresonant background dominated by the $\sigma_{_{\rm NR}}$
term defined in Eq. (\ref{eq:KKssme}). Hence, this mode is ideal
for determining the unknown parameter $\sigma_{_{\rm NR}}$ which
is given in Eq. (\ref{eq:sigma}). Time-dependent \CP violation in
neutral 3-body decay modes with fixed \CP parity was first
discussed by Gershon and Hazumi \cite{Gershon}.

Results for the decay rates and \CP asymmetries in $\ov B {}^0\to
K^+ K^- K_{S(L)},\,K_S K_S K_{S(L)}$ are displayed in
Table~\ref{tab:BrKKK} and Table~\ref{tab:ASKKK}, respectively.
(For the decay amplitudes of $\ov B {}^0\to K_S K_S K_{S(L)}$,
see \cite{CCSKKK} for details.) The mixing-induced \CP violations
are defined by
 \be \label{eq:A&S}
 S_{KKK,CP\pm}&=&   \left. \frac{2\int
 {\rm Im}(e^{-2i\beta} A_{CP\pm} \bar A^*_{CP\pm}) ds_{12} ds_{23}}{\int
 |A_{CP\pm}|^2ds_{12} ds_{23}+\int
 |\bar A_{CP\pm}|^2ds_{12} ds_{23}},\right.
 \non\\
 S_{KKK}&=&  \left. \frac{2\int
 {\rm Im}(e^{-2i\beta} A \bar A^*) ds_{12} ds_{23}}{\int
 |A|^2ds_{12} ds_{23}+\int
 |\bar A|^2ds_{12} ds_{23}}\right. 
 \non\\
 &=&f_+\,S_{KKK,CP+}+(1-f_+)\,S_{KKK,CP-},
 \en
where $A$ is the decay amplitude of $\ov B^0\to K^+K^-K_{S(L)}$ or
$K_SK_SK_{S(L)}$ and $\bar A$ is the conjugated $B^0$ decay
amplitude, and $f_+$ is the \CP even fraction defined by
 \be \label{eq:f+}
 f_+\equiv \left.{\Gamma_{CP+}+\ov \Gamma_{CP+} \over
 \Gamma+\ov\Gamma}\right|_{\phi K_S~{\rm excluded}.}
 \en
Generally, it is more convenient to define an effective $\sin
2\beta$ via $S_f\equiv -\eta_f\sin 2\beta_{\rm eff}$ with
$\eta_f=2f_+-1$ for $K^+K^-K_S$. The predicted value of $f_+$ is
consistent with the data but it is on the higher end of the
experimental measurement because the $CP$-odd contributions from
the vector mesons $\rho,\omega,\cdots,$ are OZI suppressed and the
$CP$-odd nonresonant contribution is constrained by the
$\pi^+\pi^-\pi^-$ rate.

\begin{table}[t]
\caption{Branching ratios for $\ov B {}^0\to K^+ K^- K_S,\,K_S K_S
K_S,\,K_S K_S K_L$ decays and the fraction of \CPP-even
contribution to $\ov B^0\to K^+K^-K_S$, $f_+$. The branching ratio
of \CPP-odd $K^+K^-K_S$ with $\phi K_S$ excluded is shown in
parentheses. Results for $(K^+K^-K_L)_{CP\pm}$ are identical to
those for $(K^+K^-K_S)_{CP\mp}$. For theoretical errors, see Table
\ref{tab:KpKmK0}. Experimental results are taken from \cite{HFAG}.}
\label{tab:BrKKK}
\begin{ruledtabular}
\begin{tabular}{c r r}
Final State &${\cal B}(10^{-6})_{\rm theory}$ &${\cal B}(10^{-6})_{\rm expt}$ \\
\hline
 $K^+ K^- K_S$
       & $9.89^{+0.19+2.28+0.07}_{-0.21-1.81-0.08}$
       & $12.4\pm1.2$ \\
 $(K^+ K^- K_S)_{CP+}$
       & $8.33^{+0.10+1.82+0.05}_{-0.12-1.49-0.06}$
       &  \\
 $(K^+ K^- K_S)_{CP-}$
       & $1.57^{+0.09+0.46+0.02}_{-0.10-0.32-0.02}$
       &  \\
       & $(0.14^{+0.06+0.14+0.01}_{-0.06-0.06-0.01})$
       &  \\
 $K_S K_S K_S$
       & input
       & $6.2\pm0.9$ \\
 $K_S K_S K_L$
       & $7.63^{+0.01+1.37+0.03}_{-0.01-1.19-0.03}$
       & $<14$
       \\
 \hline
& $f_+^{\rm theory}$  &$f_+^{\rm expt}$ \\
\hline $K^+K^-K_S$
           & $0.98^{+0.01+0.01+0.00}_{-0.01-0.02-0.00}$
           & $0.91\pm0.07$
           \\
\hline
&$f_-^{\rm theory}$  \\
\hline $K^+K^-K_L$
          & $0.98^{+0.01+0.01+0.00}_{-0.01-0.02-0.00}$
 \end{tabular}
\end{ruledtabular}
\end{table}

\begin{table}[t]
\caption{Mixing-induced and direct \CP asymmetries $\sin
2\beta_{\rm eff}$ (top) and $A_f$ (in $\%$, bottom), respectively,
in $\ov B^0\to K^+K^-K_S$ and $K_SK_SK_S$ decays. Experimental
results for $K^+K^-K_S$ and $K^+K^-K_L$ modes are obtained from the
data of $\ov B^0\to K^+K^-\ov K^0$. Results for
$(K^+K^-K_L)_{CP\pm}$ are identical to those for
$(K^+K^-K_S)_{CP\mp}$. For theoretical errors, see Table
\ref{tab:KpKmK0}. Experimental results are taken from \cite{HFAG}.
} \label{tab:ASKKK}
\begin{ruledtabular}
\begin{tabular}{l r r}
 Final state & $\sin 2\beta_{\rm eff}$  & Expt.  \\
 \hline
 $(K^+K^-K_S)_{\phi K_S~{\rm excluded}}$
            & $0.728^{+0.001+0.002+0.009}_{-0.002-0.001-0.020}$
            & $0.73\pm0.10$
            \\
 $(K^+K^-K_S)_{CP+}$
            & $0.732^{+0.003+0.006+0.009}_{-0.004-0.004-0.020}$
            &
            \\
 $(K^+K^-K_L)_{\phi K_L~{\rm excluded}}$
            & $0.728^{+0.001+0.002+0.009}_{-0.002-0.001-0.020}$
            & $0.73\pm0.10$
            \\
 $K_SK_SK_S$
            & $0.719^{+0.000+0.000+0.008}_{-0.000-0.000-0.019}$
            & $0.58\pm0.20$
            \\
 $K_SK_SK_L$
            & $0.718^{+0.000+0.000+0.008}_{-0.000-0.000-0.019}$
            &
            \\
 \hline
  &$A_f(\%)$  &Expt. \\
 \hline
 $(K^+K^-K_S)_{\phi K_S~{\rm excluded}}$
            & $-4.63^{+1.35+0.53+0.40}_{-1.01-0.54-0.34}$
            & $-7\pm8$
            \\
 $(K^+K^-K_S)_{CP+}$
            & $-4.86^{+1.43+0.52+0.42}_{-1.09-0.55-0.35}$
            &
            \\
 $(K^+K^-K_L)_{\phi K_L~{\rm excluded}}$
            & $-4.63^{+1.35+0.53+0.40}_{-1.01-0.54-0.34}$
            & $-7\pm8$
            \\
 $K_SK_SK_S$
            & $0.69^{+0.01+0.01+0.05}_{-0.01-0.01-0.06}$
            & $14\pm15$
            \\
 $K_SK_SK_L$
            & $0.77^{+0.01+0.01+0.05}_{-0.01-0.03-0.07}$
            &
            \\
 \end{tabular}
\end{ruledtabular}
\end{table}

The deviation of the mixing-induced \CP asymmetry in $B^0\to
K^+K^-K_S$ and $K_SK_SK_S$ from that measured in $B\to \phi_{c\bar
c}K_S$, i.e. $\sin 2 \beta_{\phi_{c\bar c}K_S}=0.681\pm0.025$
\cite{HFAG}, namely, $\Delta \sin 2\beta_{\rm eff}\equiv \sin
2\beta_{\rm eff}-\sin 2 \beta_{\phi_{c\bar c}K_S}$, is calculated
from Table \ref{tab:ASKKK} to be
\begin{eqnarray}
 \label{eq:DeltaS}
 \Delta \sin 2\beta_{K^+K^-K_S}&=&0.047^{+0.028}_{-0.033}\,, \non \\
 \Delta \sin 2\beta_{K_SK_SK_S}&=&0.038^{+0.027}_{-0.032}\,.
 \end{eqnarray}
The corresponding experimental values are $0.049\pm0.10$ and
$-0.101\pm0.20$, respectively.  Due to the presence of
color-allowed tree contributions in $\ov B^0\to K^+K^-K_{S}$, it is
naively expected that this penguin-dominated mode is subject to a
potentially significant tree pollution and hence $\Delta \sin
2\beta_{\rm eff}$ can be as large as ${\cal O}(10\%)$. However, our
calculation indicates the deviation of the mixing-induced \CP
asymmetry in $\ov B^0\to K^+K^-K_{S}$ from that measured in $\ov
B^0\to \phi_{c\bar c}K_S$ is very similar to that of the
$K_SK_SK_S$ mode as the tree pollution effect in the former is
somewhat washed out. Nevertheless, direct $CP$ asymmetry of the
former, being of order $-4\%$, is more prominent than the
latter.\footnote{ In our previous work \cite{CCSKKK},  $\Delta \sin
2\beta_{\rm eff}$ is found to be
  \be
 \label{eq:DeltaS}
 \Delta \sin 2\beta_{K^+K^-K_S}=0.06^{+0.09}_{-0.04}\,,\qquad
 \Delta \sin
 2\beta_{K_SK_SK_S}=0.06^{+0.03}_{-0.04}\,, \non
 \en
for $\sin 2\beta_{J/\psi K_S}=0.687\pm0.032$, while direct \CP
asymmetry is less than 1\% in both modes. Note that due to an
oversight the experimental error bars were not included in our
previous paper for the theoretical calculation of $\Delta \sin
2\beta_{\rm eff}$.}

\subsection{$B^-\to KKK$ decays}
The $B^-\to K^+K^-K^-$ decay amplitude has a similar expression as
Eq. (\ref{eq:AKpKmK0}) except that one also needs to add the
contributions from the interchange $s_{23}\to s_{12}$ and put a
factor of 1/2 in the decay rate to account for the identical
particle effect.

Branching ratios of resonant and nonresonant contributions to
$B^-\to K^+K^-K^-$ are shown in Table \ref{tab:KpKmKm}. It is
clear that the predicted rates of resonant and nonresonant
components are consistent with the data except for the broad
scalar resonance $X_0(1550)$. Both BaBar and Belle have seen a
large fraction from $X_0(1550)$, $(121\pm19\pm6)\%$ by BaBar
\cite{BaBarKpKpKm} and $(63.4\pm6.9)\%$ by Belle
\cite{BelleKpKpKm}, \footnote{Belle \cite{BelleKpKpKm} actually
found two solutions for the fraction of $X_0(1550)K^-$:
$(63.4\pm6.9)\%$ and $(8.21\pm1.94)\%$. The first solution is
preferred by Belle.}
while our prediction is similar to that in $\ov B^0\to K^+K^-\ov
K^0$. It is not clear why there is a huge disparity between
$B^-\to K^+K^-K^-$ and $\ov B^0\to K^+K^-\ov K^0$ as far as the
$X_0(1550)$ contribution is concerned. Obviously, a refined
measurement of the $X_0(1550)$ contribution to the $K^+K^-K^-$
mode is urgently needed in order to clarify this issue. Our result
for the nonresonant contribution is in good agreement with Belle,
but disagrees with BaBar. Notice that Belle did not see the scalar
resonance $f_0(980)$ as Belle employed the E791 result \cite{E791}
for $g^{f_0\to K\bar K}$ which is smaller than $g^{f_0\to\pi\pi}$.
In contrast to E791, the ratio $g^{f_0\to K\bar K}/g^{f_0\to
\pi\pi}$ is measured to be larger than 4 in the existing $e^+e^-$
experiments \cite{BES,epem}

\begin{table}[h]
\caption{Branching ratios (in units of $10^{-6}$) of resonant and
nonresonant (NR) contributions to $B^-\to K^+K^-K^-$. For
theoretical errors, see Table \ref{tab:KpKmK0}.}
\begin{ruledtabular} \label{tab:KpKmKm}
\begin{tabular}{l l l l}
 Decay mode~~ & BaBar \cite{BaBarKpKpKm} & Belle \cite{BelleKpKpKm} & Theory  \\ \hline
 $\phi K^-$ & $4.14\pm0.32\pm0.33$ &
$4.72\pm0.45\pm0.35^{+0.39}_{-0.22}$ & $2.9^{+0.0+0.5+0.0}_{-0.0-0.5-0.0}$ \\
 $f_0(980)K^-$ & $6.5\pm2.5\pm1.6$ & $<2.9$ & $7.0^{+0.0+0.4+0.1}_{-0.0-0.7-0.1}$ \\
 $X_0(1550)K^-$ & $43\pm6\pm3$ &  & $1.1^{+0.0+0.2+0.0}_{-0.0-0.2-0.0}$ \\
 $f_0(1710)K^-$ & $1.7\pm1.0\pm0.3$ & & \\
 NR & $50\pm6\pm4$ & $24.0\pm1.5\pm1.8^{+1.9}_{-5.7}$ & $25.3^{+0.9+4.8+0.3}_{-1.0-4.4-0.3}$ \\
\hline
 Total & $35.2\pm0.9\pm1.6$ & $32.1\pm1.3\pm2.4$ &  $25.5^{+0.5+4.4+0.2}_{-0.6-4.1-0.2}$ \\
\end{tabular}
\end{ruledtabular}
\end{table}

We next turn to the decay $B^-\to K^-K_SK_S$. Following
\cite{Gronau3body}, let us consider the symmetric state of $K^0\ov
K^0$
 \be
 |K^0\ov K^0\ra_{\rm sym} &\equiv& \left[|K^0(p_1)\ov K^0(p_2)\ra +|\ov
 K^0(p_1)K^0(p_2)\ra\right]/\sqrt{2} \non \\
 &=&
 \left[|K_S(p_1)K_S(p_2)\ra-|K_L(p_1)K_L(p_2)\ra\right]/\sqrt{2}.
 \en
Hence,
 \be
 \B(B^-\to K^-K_SK_S) &=& {1\over 2}[\B(B^-\to K^-K_SK_S)+\B(B^-\to
 K^-K_LK_L)]
\non \\
 &=& {1\over 2}\B(B^-\to K^-(K^0\ov K^0)_{\rm sym}).
 \en
The factorizable amplitude of $B^-\to K^-K^0\ov K^0$ is given by
Eq. (\ref{eq:AKmK0K0}). Just as other $KKK$ modes, this decay is
also expected to be dominated by the nonresonant contribution (see
Table \ref{tab:KmKsKs}). The calculated total rate is in good
agreement with experiment. Just as the pure penguin mode
$K_SK_SK_S$, the decay $B^-\to K^-K_SK_S$ also can be used to
constrain the nonresonant parameter $\sigma_{_{\rm NR}}$.

As pointed out in \cite{Gronau3body}, isospin symmetry implies the
relation
 \be \label{eq:isospin}
 A(B^-\to K^-K^0\ov K^0)=-A(\ov B^0\to \ov K^0K^+K^-).
 \en
This leads to
 \be
 \B(B^-\to K^-(K^0\ov K^0)_{\rm sym})={\tau(B^-)\over\tau(B^0)}
 \B(\ov B^0\to K^+K^-\ov K^0)_{\phi K ~{\rm excluded}}.
 \en
Experimentally, this relation is well satisfied:
LHS=$(23.0\pm2.6)\times 10^{-6}$ and RHS=$(22.1\pm2.1)\times
10^{-6}$. Hence, the isospin relation Eq. (\ref{eq:isospin}) is
well respected.

\begin{table}[h]
\caption{Branching ratios (in units of $10^{-6}$) of resonant and
nonresonant (NR) contributions to $B^-\to K^-K_SK_S$. For
theoretical errors, see Table \ref{tab:KpKmK0}.}
\begin{center} \label{tab:KmKsKs}
\begin{tabular}{|c| c c c | c |} \hline
~~Decay mode~~ & ~~$f_0(980)K^-~~$ & ~~$X_0(1550)K^-$~~ & ~~NR~~ & ~~total~~ \\
\hline Theory & ~~$5.2^{+0.0+0.3+0.1}_{-0.0-0.5-0.1}$ &
$0.92^{+0.00+0.16+0.00}_{-0.00-0.15-0.00}$
& $12.4^{+0.2+2.1+0.1}_{-0.3-2.0-0.1}$~~ & ~~$12.2^{+0.0+1.5+0.0}_{-0.0-1.7-0.0}$ \\
Expt. & & & & ~~$11.5\pm1.3$~~ \\
 \hline
\end{tabular}
\end{center}
\end{table}

\section{$B\to K\pi\pi$ decays}
In this section we shall consider five $B\to K\pi\pi$ decays,
namely, $B^-\to K^-\pi^+\pi^-$, $\ov K^0\pi^-\pi^0$, $\ov B^0\to
K^-\pi^+\pi^0$, $\ov K^0\pi^+\pi^-$ and $\ov K^0\pi^0\pi^0$. They
are dominated by $b\to s$ penguin transition and consist of three
decay processes: (i) the current-induced process, $\la B\to \pi\pi
\ra\times \la 0\to K\ra$, (ii) the transition processes, $\la B\to
\pi\ra\times \la 0\to \pi K\ra$, and $\la B\to K\ra\times \la 0\to
\pi\pi\ra$, and (iii) the annihilation process $\la B\to
0\ra\times \la 0\to K\pi\pi\ra$.

The factorizable amplitudes for  $B^-\to K^-\pi^+\pi^-$, $\ov
K^0\pi^-\pi^0$, $\ov B^0\to K^-\pi^+\pi^0$, $\ov K^0\pi^+\pi^-$
and $\ov K^0\pi^0\pi^0$ are given in Eqs.
(\ref{eq:AKmpippim}-\ref{eq:AK0pi0pi0}), respectively. All five
channels have the three-body matrix element $\la \pi \pi|(\bar q
b)_{V-A}|B\ra$ which has the similar expression as Eqs.
(\ref{eq:AHMChPT}) and (\ref{eq:r&omega}) except that the pole
$B_s^*$ is replaced by $B^*$ and the kaon is replaced by the pion.
However, there are additional resonant contributions to this
three-body matrix element due to the intermediate vector $\rho$
and scalar $f_0$ mesons
 \be
 \la \pi^+(p_2)\pi^-(p_3)|(\bar ub)_\vma|B^-\ra^R &=& \sum_i
{g^{\rho_i^0\to\pi^+\pi^-}\over
m_{\rho_i}^2-s_{23}-im_{\rho_i}\Gamma_{\rho_i}}\sum_{\rm
pol}\vp^*\cdot
(p_2-p_3)\la \rho_i^0|(\bar ub)_\vma|B^-\ra \non \\
&+& \sum_i{g^{{f_0}_i\to\pi^+\pi^-}\over
m_{{f_0}_i}^2-s_{23}-im_{{f_0}_i}\Gamma_{{f_0}_i}}\la
{f_0}_i|(\bar ub)_\vma|B^-\ra,
 \en
where $\rho_i$ denote generic $\rho$-type vector mesons, e.g.
$\rho=\rho(770),\rho(1450),\rho(1700),\cdots$. Applying Eqs.
(\ref{eq:decayc}) and (\ref{eq:FF}) we are led to
 \be
 && \la \pi^+(p_2)\pi^-(p_3)|(\bar ub)\vma|B^-\ra^R ~\la K^-(p_1)|(\bar
su)_\vma|0\ra \non \\
&=& \sum _i{f_K\over 2}\,{g^{\rho_i^0\to\pi^+\pi^-}\over
m_{\rho_i^2}-s_{23}-im_{\rho_i}\Gamma_{\rho_i}}(s_{12}-s_{13})\Big[
(m_B+m_{\rho_i})A_1^{B\rho_i}(q^2) \non \\ &-&
{A_2^{B\rho_i}(q^2)\over m_B+m_{\rho_i}}
(s_{12}+s_{13}-3m_\pi^2)-2m_{\rho_i}[A_3^{B\rho_i}(q^2)-A_0^{B\rho_i}(q^2)]\Big] \non\\
&+& \sum _i{f_K \,g^{{f_0}_i\to\pi^+\pi^-}\over
m_{{f_0}_i}^2-s_{23}-im_{{f_0}_i}\Gamma_{{f_0}_i}}(m_B^2-m_{{f_0}_i}^2)F_0^{Bf_0^u}(q^2).
 \en
Likewise, the 3-body matrix element $\la K^-\pi^+|(\bar
sb)_\vma|\ov B^0\ra$ appearing in $\ov B^0\to K^-\pi^+\pi^0$ also
receives the following resonant contributions
 \be
 \la K^-(p_1)\pi^+(p_2)|(\bar sb)_\vma|\ov B^0\ra^R &=& \sum_i
{g^{K_i^*\to K^-\pi^+}\over
m_{K_i^*}^2-s_{12}-im_{K_i^*}\Gamma_{K_i^*}}\sum_{\rm
pol}\vp^*\cdot
(p_1-p_2)\la \ov K^{*0}_i|(\bar sb)_\vma|\ov B^0\ra, \non \\
 \en
with $K_i^*=K^*(892), K^*(1410),K^*(1680),\cdots$.

For the two-body matrix elements $\la \pi^+K^-|(\bar
sd)_{V-A}|0\ra$, $\la \pi^+\pi^-|(\bar uu)_{V-A}|0\ra$ and $\la
\pi^+\pi^-|\bar ss|0\ra$, we note that
 \be
 \la K^-(p_1)\pi^+(p_2)|(\bar sd)_\vma|0\ra &=& \la \pi^+(p_2)|(\bar
 sd)_\vma|K^+(-p_1)\ra = (p_1-p_2)_\mu F_1^{K\pi}(s_{12}) \non \\
 &+& {m_K^2-m_\pi^2\over
 s_{12}}(p_1+p_2)_\mu\left[-F_1^{K\pi}(s_{12})+F_0^{K\pi}(s_{12})\right],
 \en
where we have taken into account the sign flip arising from
interchanging the operators $s\leftrightarrow d$. Hence,
 \be \label{eq:KpiBpi}
&& \la K^-(p_1)\pi^+(p_2)|(\bar s d)_{V-A}|0\ra
\la \pi^-(p_3)|(\bar d b)_{V-A}|B^-\ra  \non\\
&=&F_1^{B\pi}(s_{12})F_1^{K\pi}(s_{12})\left[s_{23}-s_{13}-{(m_B^2-m_\pi^2)(m_K^2-m_\pi^2)
\over s_{12}}\right] \non
\\ &+&  F_0^{B\pi}(s_{12})F_0^{K\pi}(s_{12}){(m_B^2-m_\pi^2)(m_K^2-m_\pi^2)
\over s_{12}}.
 \en
However, the form factor $F_1$ also receives resonant
contributions
 \be
 \sum_i \left(A^\mu_{K_i^*\pi K}\,{1\over
 m_{K_i^*}^2-s_{12}-im_{K_i^*}\Gamma_{K_i^*}}\,m_{K_i^*}f_{K_i^*}+{g^{K^*_{0i}\to K\pi}\over
 m_{K_{0i}^*}^2-s_{12}-im_{K_{0i}^*}\Gamma_{K_{0i}^*}}\,f_{K_{0i}^*}(p_1-p_2)_\mu\right),
 \en
with
 \be
 \vp_\mu^* A^\mu_{K^*\pi K}=\la K^-(p_1)\pi^+(p_2)|K^*\ra=g^{K^*\to \pi K}\,\vp^*\cdot
 (p_1-p_2),
 \en
where ${K^*_0}_i=K^*_0(1430),\cdots$. Hence, the resonant
contributions to the form factor $F_1^{K\pi}$ are
 \be \label{eq:F1Kpi}
 F^{K\pi}_{1,R}(s)=\sum_i\left({m_{K_i^*}f_{K_i^*}g^{K_i^*\to K\pi}\over
 m_{K_i^*}^2-s-im_{K_i^*}\Gamma_{K_i^*}}+{f_{K_{0i}^*}g^{K^*_{0i}\to K\pi}\over
 m_{K_{0i}^*}^2-s-im_{K_{0i}^*}\Gamma_{K_{0i}^*}}\right).
 \en
In principle, the weak vector form factor $F^{\pi^+\pi^-}$ defined
by
 \be
 \la \pi^+(p_{\pi^+})\pi^-(p_{\pi^-})|\bar u\gamma_\mu u|0\ra
 &=& (p_{\pi^+}-p_{\pi^-})_\mu F^{\pi^+\pi^-},
 \en
can be related to the time-like pion electromagnetic form factors.
However, unlike the kaon case, the time-like e.m. form factors of
the pions are not well measured enough allowing us to determine
the resonant and nonresonant parts. Therefore, we shall only
consider the resonant part which has the expression
 \be
 F^{\pi\pi}_R(s)=\sum_i{m_{\rho_i}f_{\rho_i}g^{\rho_i\to \pi\pi}\over
 m_{\rho_i}^2-s-im_{\rho_i}\Gamma_{\rho_i}}.
 \en

Following Eq. (\ref{eq:KKssme}), the relevant matrix elements of
scalar densities read
 \be
 \la \pi^+(p_2) \pi^-(p_3)|\bar s s|0\ra
 =\sum_{i}\frac{m_{{f_0}_i} \bar f^s_{{f_0}_i} g^{{f_0}_i\to \pi^+\pi^-}}{m_{{f_0}_i}^2-s_{23}-i
 m_{{f_0}_i}\Gamma_{{f_0}_i}}+ \la \pi^+(p_2) \pi^-(p_3)|\bar s
 s|0\ra^{NR},
 \en
and
 \be
 \la K^-(p_1) \pi^+(p_2)|\bar s d|0\ra
 =\sum_{i}\frac{m_{{K^*_0}_i} \bar f_{{K^*_0}_i} g^{{K^*_0}_i\to K^-\pi^+}}{m_{{K^*_0}_i}^2-s_{12}-i
 m_{{K^*_0}_i}\Gamma_{{K^*_0}_i}}+\la K^-(p_1) \pi^+(p_2)|\bar s
 d|0\ra^{NR}.
 \en
Note that for the scalar meson, the decay constants $f_S$ and
$\bar f_S$ are defined in Eq. (\ref{eq:decayc}) and they are
related via Eq. (\ref{eq:EOM}). The nonresonant contribution $\la
\pi^+(p_2) \pi^-(p_3)|\bar s s|0\ra^{NR}$ vanishes under the OZI
rule, while under SU(3) symmetry\footnote{The matrix elements of
scalar densities can be generally decomposed into $D$-, $F$- and
$S$(singlet)-type components. Assuming that the singlet component
is OZI suppressed, SU(3) symmetry leads to, for example, the
relation $\la K\pi|\bar sq|0\ra^{NR}=\la K\bar K|\bar
ss|0\ra^{NR}$.}
 \be
 \la K^-(p_1) \pi^+(p_2)|\bar sd|0\ra^{NR}=\la K^+(p_1)K^-(p_2)|\bar
 ss|0\ra^{NR}=f_s^{NR}(s_{12}),
 \en
with the expression of $f_s^{NR}$ given in Eq. (\ref{eq:KKssme}).

It is known that in the narrow width approximation, the 3-body
decay rate obeys the factorization relation
 \be \label{eq:fact}
 \Gamma(B\to RP\to P_1P_2P)=\Gamma(B\to RP)\B(R\to P_1P_2),
 \en
with $R$ being a resonance. This means that the amplitudes $A(B\to
RP\to P_1P_2P)$ and $A(B\to RP)$ should have the same expressions
apart from some factors. Hence, using the known results for
quasi-two-body decay amplitude $A(B\to RP)$, one can have a cross
check on the three-body decay amplitude of $B\to RP\to P_1P_2P$.
For example, from Eq. (\ref{eq:AKmpippi0}) we obtain the
factorizable amplitude  $A(\ov B^0\to K_0^{*0}(1430)\pi^0;
K^{*0}_0(1430)\to K^-\pi^+)$ as
 \be
&& \la K^-(p_1)\pi^+(p_2)\pi^0(p_3)|T_p|\ov
B^0\ra_{K_0^{*0}(1430)} = \non
\\ && {1\over\sqrt{2}}{g^{K_0^{*0}(1430)\to K^-\pi^+}\over
m_{K_0^*}^2-s_{12}-im_{K_0^*}\Gamma_{K_0^*}}\Bigg\{ \left(
-a_4^p+r_\chi^{K^*_0}a_6^p
 +{1\over 2}(a_{10}^p-r_\chi^{K^*_0}a_8^p) \right)
 f_{K_0^*}F_0^{B\pi}(m_{K_0^*}^2)(m_B^2-m_\pi^2) \non \\
 && -\left[a_2\delta_{pu}+{3\over
 2}(a_9-a_7)\right]f_\pi F_0^{BK^*_0}(m_\pi^2)(m_B^2-m_{K_0^*}^2)
 \Bigg\},
 \en
where
 \be
  r^{K^*_0}_\chi(\mu)={2m_{K_0^*}^2\over
 m_b(\mu)(m_s(\mu)-m_q(\mu))}.
 \en
The expression inside $\{\cdots\}$ is indeed the amplitude of $\ov
B^0\to K_0^{*0}(1430)\pi^0$ given in Eq. (A6) of \cite{CCY}.

The strong coupling constants such as $g^{\rho\to\pi^+\pi^-}$ and
$g^{f_0(980)\to\pi^+\pi^-}$ are determined from the measured
partial widths through the relations
 \be
 \Gamma_S={p_c\over 8\pi m_S^2}g_{S\to P_1P_2}^2,\qquad
 \Gamma_V={2\over 3}\,{p_c^3\over 4\pi m_V^2}g_{V\to P_1P_2}^2,
 \en
for scalar and vector mesons, respectively, where $p_c$ is the
c.m. momentum. The numerical results are
 \be \label{eq:g}
 && g^{\rho\to\pi^+\pi^-}=6.0, \qquad\qquad\qquad\quad
\quad g^{K^*\to K^+\pi^-}=4.59,\non \\
&& g^{f_0(980)\to\pi^+\pi^-}=1.33^{+0.29}_{-0.26}\,{\rm GeV},
\quad g^{K_0^*\to K^+\pi^-}=3.84\,{\rm GeV}.
 \en
In determining the coupling of $f_0\to\pi^+\pi^-$, we have used
the partial width
 \be
 \Gamma(f_0(980)\to\pi^+\pi^-)=(34.2^{+13.9+8.8}_{-11.8-2.5})\,{\rm
 MeV}
 \en
measured by Belle \cite{Bellef0}. The momentum dependence of the
weak form factor $F^{K\pi}(q^2)$ is parametrized as
 \be \label{Kpi}
 F^{K\pi}(q^2)=\,{F^{K\pi}(0)\over 1-q^2/{\Lambda_\chi}^2+i\Gamma_R/{\Lambda_\chi}},
 \en
where $\Lambda_\chi\approx 830$ MeV is the chiral-symmetry
breaking scale \cite{Cheng88} and $\Gamma_R$ is the width of the
relevant resonance, which is taken to be 200 MeV
\cite{Cheng:2002qu}.

The results of the calculation are summarized in Tables
\ref{tab:Kpipi}-\ref{tab:K0pi0pi0}. We see that except for
$f_0(980)K$,  the predicted rates for $K^{*}\pi$,
$K^{*}_0(1430)\pi$ and $\rho K$ are smaller than the data. Indeed,
the predictions based on QCD factorization for these decays are
also generally smaller than experiment by a factor of 2$\sim$5.
This will be discussed in more details in Sec. VI.

\begin{table}[h]
\caption{Branching ratios (in units of $10^{-6}$) of resonant and
nonresonant (NR) contributions to $B^-\to K^-\pi^+\pi^-$. For
theoretical errors, see Table \ref{tab:KpKmK0}.}
\begin{ruledtabular}
\begin{tabular}{l l l l} \label{tab:Kpipi}
Decay mode~~ & BaBar \cite{BaBarKpipi} & Belle \cite{BelleKpipi} &
Theory
\\ \hline
 $\ov K^{*0}\pi^-$ & $9.04\pm0.77\pm0.53^{+0.21}_{-0.37}$
 & $6.45\pm0.43\pm0.48^{+0.25}_{-0.35}$ &  $3.0^{+0.0+0.8+0.0}_{-0.0-0.7-0.0}$ \\
 $\ov K^{*0}_0(1430)\pi^-$ & $34.4\pm1.7\pm1.8^{+0.1}_{-1.4}$ &
$32.0\pm1.0\pm2.4^{+1.1}_{-1.9}$ & $10.5^{+0.0+3.2+0.0}_{-0.0-2.7-0.1}$  \\
 $\rho^0K^-$ & $5.08\pm0.78\pm0.39^{+0.22}_{-0.66}$ &
$3.89\pm0.47\pm0.29^{+0.32}_{-0.29}$ & $1.3^{+0.0+1.9+0.1}_{-0.0-0.7-0.1}$ \\
 $f_0(980)K^-$ & $9.30\pm0.98\pm0.51^{+0.27}_{-0.72}$
 & $8.78\pm0.82\pm0.65^{+0.55}_{-1.64}$ & $7.7^{+0.0+0.4+0.1}_{-0.0-0.8-0.1}$  \\
NR & $2.87\pm0.65\pm0.43^{+0.63}_{-0.25}$ &
$16.9\pm1.3\pm1.3^{+1.1}_{-0.9}$ & $18.7^{+0.5+11.0+0.2}_{-0.6-~6.3-0.2}$ \\
\hline
 Total & $64.4\pm2.5\pm4.6$ & $48.8\pm1.1\pm3.6$ & $45.0^{+0.3+16.4+0.1}_{-0.4-10.5-0.1}$ \\
\end{tabular}
\end{ruledtabular}
\end{table}

\begin{table}[h]
\caption{Same as Table \ref{tab:Kpipi} except for the decay $B^-\to
\ov K^0\pi^-\pi^0$. }
\begin{ruledtabular} \label{tab:K0pimpi0}
\begin{tabular}{l l | l l }
 Decay mode~~~~~~~~~~~~~ & Theory~~~~~~~~~~~~~~~~~~~~~  & Decay mode & Theory \\ \hline
 $K^{*-}\pi^0$ & $1.5^{+0.0+0.3+0.2}_{-0.0-0.3-0.2}$ & $\ov
 K^{*0}\pi^-$ & $1.5^{+0.0+0.4+0.0}_{-0.0-0.3-0.0}$ \\
 $K^{*-}_0(1430)\pi^0$ & $5.5^{+0.0+1.6+0.1}_{-0.0-1.4-0.1}$ & $\ov
 K^{*0}_0(1430)\pi^-$ & $5.2^{+0.0+1.6+0.0}_{-0.0-1.4-0.0}$ \\
 $\rho^-\ov K^0$ & $1.3^{+0.0+3.0+0.0}_{-0.0-0.9-0.0}$ & NR &
 $10.0^{+0.2+7.1+0.0}_{-0.2-3.7-0.0}$ \\ \hline
 Total & $27.0^{+0.3+15.4+0.2}_{-0.2-~8.8-0.2}$ & &
\end{tabular}
\end{ruledtabular}
\end{table}

\begin{table}[h]
\caption{Same as Table \ref{tab:Kpipi} except for the decay $\ov
B^0\to \ov K^0\pi^+\pi^-$. }
\begin{ruledtabular} \label{tab:K0pipi}
\begin{tabular}{l l l }
 Decay mode~~ &  Belle \cite{BelleK0pipi} & Theory  \\ \hline
 $K^{*-}\pi^+$ & $5.6\pm0.7\pm0.5^{+0.4}_{-0.3}$ &  $2.1^{+0.0+0.5+0.3}_{-0.0-0.5-0.3}$ \\
 $K^{*-}_0(1430)\pi^+$  & $30.8\pm2.4\pm2.4^{+0.8}_{-3.0}$ & $10.1^{+0.0+2.9+0.1}_{-0.0-2.5-0.2}$ \\
 $\rho^0\ov K^0$  & $6.1\pm1.0\pm0.5^{+1.0}_{-1.1}$ & $2.0^{+0.0+1.9+0.1}_{-0.0-0.9-0.1}$ \\
 $f_0(980)\ov K^0$ & $7.6\pm1.7\pm0.7^{+0.5}_{-0.7}$ & $7.7^{+0.0+0.4+0.0}_{-0.0-0.7-0.0}$ \\
 NR  & $19.9\pm2.5\pm1.6^{+0.7}_{-1.2}$ & $15.6^{+0.1+8.3+0.0}_{-0.1-4.9-0.0}$ \\ \hline
 Total  & $47.5\pm2.4\pm3.7$ & $42.0^{+0.3+15.7+0.0}_{-0.2-10.8-0.0}$ \\
\end{tabular}
\end{ruledtabular}
\end{table}

\begin{table}[h]
\caption{Branching ratios (in units of $10^{-6}$) of resonant and
nonresonant (NR) contributions to $\ov B^0\to K^-\pi^+\pi^0$. Note
that the branching ratios for $K^{*-}\pi^+$ and $\ov K^{*0}\pi^0$
given in \cite{BaBarKppimpi0} and \cite{BelleKppimpi0} are their
absolute ones. We have converted them into the product branching
ratios, namely, $\B(B\to Rh)\times \B(R\to hh)$. For theoretical
errors, see Table \ref{tab:KpKmK0}.}
\begin{ruledtabular} \label{tab:Kmpippi0}
\begin{tabular}{l l l l}
 Decay mode~~ & BaBar \cite{BaBarKppimpi0} & Belle \cite{BelleKppimpi0} & Theory  \\ \hline
 $K^{*-}\pi^+$ & $3.6\pm0.8\pm0.5$  & $4.9^{+1.5+0.5+0.8}_{-1.5-0.3-0.3}$ &   $1.0^{+0.0+0.3+0.1}_{-0.0-0.3-0.1}$ \\
 $\ov K^{*0}\pi^0$ & $2.0\pm0.6\pm0.3$  & $<2.3$ &  $1.0^{+0.0+0.3+0.2}_{-0.0-0.2-0.1}$  \\
 $K^{*-}_0(1430)\pi^+$ & $11.2\pm1.5\pm3.5$ &  $5.1\pm1.5^{+0.6}_{-0.7}$ & $5.0^{+0.0+1.5+0.1}_{-0.0-1.3-0.1}$ \\
 $\ov K^{*0}_0(1430)\pi^0$ & $7.9\pm1.5\pm2.7$ &  $6.1^{+1.6+0.5}_{-1.5-0.6}$ & $4.2^{+0.0+1.4+0.0}_{-0.0-1.2-0.0}$ \\
 $\rho^+K^-$ & $8.6\pm1.4\pm1.0$ & $15.1^{+3.4+1.4+2.0}_{-3.3-1.5-2.1}$ & $2.5^{+0.0+3.6+0.2}_{-0.0-1.4-0.2}$ \\
 NR & $<4.6$ & $5.7^{+2.7+0.5}_{-2.5-0.4}<9.4$ & $9.6^{+0.3+6.6+0.0}_{-0.2-3.5-0.0}$ \\ \hline
 Total & $34.9\pm2.1\pm3.9$ & $36.6^{+4.2}_{-4.1}\pm3.0$ & $28.9^{+0.2+16.1+0.2}_{-0.2-~9.4-0.2}$ \\
\end{tabular}
\end{ruledtabular}
\end{table}

\begin{table}[h]
\caption{Same as Table \ref{tab:Kpipi} except for the decay $\ov
B^0\to \ov K^0\pi^0\pi^0$.}
\begin{center} \label{tab:K0pi0pi0}
\begin{tabular}{|c| c c c c| c |} \hline
Decay mode~~ & ~~$f_0(980)\ov K^0~~$ & ~~$\ov K^{*0}\pi^0$~~ & ~~$\ov K^{*0}_0(1430)\pi^0$~~ & ~~NR~~ & ~~Total~~ \\
\hline Theory & $3.8^{+0.0+2.0+0.0}_{-0.0-0.4-0.0}$ &
$0.55^{+0.00+0.16+0.00}_{-0.00-0.13-0.00}$ &
$2.3^{+0.0+0.8+0.0}_{-0.0-0.6-0.0}$ &
$5.3^{+0.0+1.8+0.0}_{-0.0-1.1-0.0}$ & $12.9^{+0.0+4.0+0.1}_{-0.0-3.0-0.1}$ \\
 \hline
\end{tabular}
\end{center}
\end{table}

While Belle has found a sizable fraction of order $(35\sim40)\%$
for the nonresonant signal in $K^-\pi^+\pi^-$ and $\ov
K^0\pi^+\pi^-$ modes (see Table \ref{tab:BRexpt}), BaBar reported
a small fraction of order 4.5\% in $K^-\pi^+\pi^-$. The huge
disparity between BaBar and Belle is ascribed to the different
parameterizations adopted by both groups. BaBar \cite{BaBarKpipi}
used the LASS parametrization to describe the $K\pi$ $S$-wave and
the nonresonant component by a single amplitude suggested by the
LASS collaboration to describe the scalar amplitude in elastic
$K\pi$ scattering. As commented in \cite{BelleKpipi}, while this
approach is experimentally motivated, the use of the LASS
parametrization is limited to the elastic region of $M(K\pi)\lsim
2.0$ GeV, and an additional amplitude is still required for a
satisfactory description of the data. In our calculations we have
taken into account the nonresonant contributions to the two-body
matrix elements of scalar densities, $\la K\pi|\bar sq|0\ra$.
Recall that a large nonresonant contribution from $\la K\ov K|\bar
ss|0\ra$ is needed in order to explain the observed decay rates of
$B^0\to K_SK_SK_S$ and $B^-\to K^-K_SK_S$. From Tables
\ref{tab:Kpipi}-\ref{tab:K0pi0pi0} we see that our predicted
nonresonant rates are in agreement with the Belle measurements.
The reason why the nonresonant fraction is as large as 90\% in
$KKK$ decays, but becomes only $(35\sim 40)\%$ in $K\pi\pi$
channels (see Table \ref{tab:BRexpt}) can be explained as follows.
Under SU(3) flavor symmetry, we have the relation $\la K\pi|\bar
sq|0\ra^{NR}=\la K\bar K|\bar ss|0\ra^{NR}$.
Hence, the nonresonant rates in the $K^-\pi^+\pi^-$ and $\ov
K^0\pi^+\pi^-$ modes should be similar to that in $K^+K^-\ov K^0$
or $K^+K^-K^-$. Since the $KKK$ channel receives resonant
contributions only from $\phi$ and $f_{0_i}$ mesons, while $K^*_i,
K^*_{0i},\rho_i,f_{0i}$ resonances contribute to $K\pi\pi$ modes,
this explains why the nonresonant fraction is of order 90\% in the
former and becomes of order 40\% in the latter. Note that the
predicted nonresonant contribution in the $K^-\pi^+\pi^0$ mode is
larger than the BaBar's upper bound and barely consistent with the
Belle limit. It is conceivable that the SU(3) breaking effect in
$\la K\pi|\bar sq|0\ra^{NR}$ may lead to a result consistent with
the Belle limit.

It is interesting to notice that, based on a simple fragmentation
model and SU(3) symmetry, Gronau and Rosner \cite{Gronau3body}
found the relations
 \be
 \Gamma(B^-\to K^+K^-K^-)_{\rm NR} &=& 2\Gamma(\ov B^0\to K^+K^-\ov K^0)_{\rm NR}
 =2\Gamma(B^-\to K^-\pi^+\pi^-)_{\rm NR} \non \\
 &=& 2\Gamma(\ov B^0\to \ov K^0\pi^+\pi^-)_{\rm NR}=4\Gamma(\ov B^0\to K^-\pi^+\pi^0)_{\rm
 NR}.
 \en
Again, a large nonresonant background in $K^-\pi^+\pi^-$ and $\ov
K^0\pi^+\pi^-$ is favored by this model.

Although the $\ov B^0\to K_S \pi^0\pi^0$ rate has not been
measured, its time-dependent \CP asymmetries have been studied by
BaBar \cite{BaBarK0pi0pi0} with the results
 \be
 \sin 2\beta_{\rm eff}=-0.72\pm0.71\pm0.08, \qquad A_{\rm
 CP}=-0.23\pm0.52\pm0.13\,.
 \en
Note that this mode is a \CPP-even eigenstate. We found that its
branching ratio is not so small, of order $6\times 10^{-6}$, in
spite of the presence of two neutral pions in the final state (see
Table \ref{tab:K0pi0pi0}). Theoretically, we obtain
 \be
  \sin 2\beta_{\rm eff}=0.729^{+0.000+0.001+0.009}_{-0.000-0.001-0.020}, \qquad A_{\rm
 CP}=\left(0.28^{+0.09+0.07+0.02}_{-0.06-0.06-0.02}\right)\%.
 \en

Finally, we consider the mode $K_S\pi^+\pi^-$ which is an
admixture of \CPP-even and \CPP-odd components. Results for the
decay rates and \CP asymmetries are displayed in Table
\ref{tab:Kspippim}. We see that the effective $\sin2\beta$ is of
order 0.718 and direct \CP asymmetry of order 4.9\% for
$K_S\pi^+\pi^-$.

\begin{table}[t]
\caption{Branching ratios, mixing-induced and direct \CP
asymmetries for $\ov B^0\to K_{S}\pi^+\pi^-$ decays. Results for
$(K_L\pi\pi)_{CP\pm}$ are identical to those for
$(K_S\pi\pi)_{CP\mp}$. For theoretical errors, see Table
\ref{tab:KpKmK0}.} \label{tab:Kspippim}
\begin{ruledtabular}
\begin{tabular}{l r }
 Final state & Branching ratio    \\
 \hline
 $(K_S\pi^+\pi^-)_{CP+}$
            & $13.52^{+0.02+4.03+0.01}_{-0.03-3.06-0.01}$
            \\
 $(K_S\pi^+\pi^-)_{CP-}$
            & $7.45^{+0.10+3.79+0.02}_{-0.08-2.32-0.02}$
            \\
 ~~~$f_+$      & $0.65^{+0.00+0.03+0.00}_{-0.00-0.04-0.00}$
            \\
 \hline
 Final state & $\sin 2\beta_{\rm eff}$    \\
 \hline
 $(K_S\pi^+\pi^-)_{CP+}$
            & $0.693^{+0.000+0.003+0.003}_{-0.000-0.002-0.014}$
            \\
 $(K_S\pi^+\pi^-)_{\rm full}$
             & $0.718^{+0.001+0.017+0.008}_{-0.001-0.007-0.018}$
            \\
 \hline
 Final state &$A_f(\%)$  \\
 \hline
 $(K_S\pi^+\pi^-)_{CP+}$
            & $4.27^{+0.00+0.19+0.28}_{-0.00-0.12-0.35}$
            \\
 $(K_S\pi^+\pi^-)_{\rm full}$
             & $4.94^{+0.03+0.03+0.32}_{-0.02-0.05-0.40}$
            \\
\end{tabular}
\end{ruledtabular}
\end{table}

\begin{table}[h]
\caption{Same as Table \ref{tab:Kpipi} except for the decay $B^-\to
K^+K^-\pi^-$.}
\begin{center} \label{tab:KpKmpim}
\begin{tabular}{|c| c c c c| c |} \hline
Decay mode & ~~$f_0(980)\pi^-~~$ & ~~$K^{*0}K^-$~~ & ~~$K^{*0}_0(1430)K^-$~~ & ~~NR~~ & ~~Total~~ \\
\hline Theory & $0.50^{+0.00+0.06+0.02}_{-0.00-0.04-0.02}$ &
$0.23^{+0.00+0.04+0.02}_{-0.00-0.04-0.02}$
& $0.82^{+0.00+0.18+0.09}_{-0.00-0.16-0.08}$ & $1.8^{+0.5+0.4+0.2}_{-0.5-0.2-0.2}$ & $4.0^{+0.5+0.7+0.3}_{-0.6-0.5-0.3}$ \\
Expt. & & & & & $<6.3~({\rm BaBar})$\cite{BaBar2003} \\
 & & & & & ~$<13~({\rm Belle})$ \cite{Belle2004}~~ \\
 \hline
\end{tabular}
\end{center}
\end{table}

\section{$B\to KK\pi$ decays}
We now turn to the three-body decay modes dominated by $b\to u$
tree and $b\to d$ penguin transitions, namely, $KK\pi$ and
$\pi\pi\pi$. We first consider the decay $B^-\to K^+K^-\pi^-$
whose factorizable amplitude is given by Eq. (\ref{eq:AKpKmpim}).
Note that we have included the matrix element $\la K^+K^-|\bar
dd|0\ra$. Although its nonresonant contribution vanishes as $K^+$
and $K^-$ do not contain the valence $d$ or $\bar d$ quark, this
matrix element does receive a contribution from the scalar $f_0$
pole
 \be
 \la K^+(p_2)K^-(p_3)|\bar dd|0\ra^R
 =\sum_{i}\frac{m_{{f_0}_i} \bar f_{{f_0}_i}^d g^{{f_0}_i\to \pi^+\pi^-}}{m_{{f_0}_i}^2-s_{23}-i
  m_{{f_0}_i}\Gamma_{{f_0}_i}},
 \en
where $\la f_0|\bar dd|0\ra=m_{f_0}\bar f_{f_0}^d$. In the 2-quark
model for $f_0(980)$, $\bar f_{f_0(980)}^d=\bar
f_{f_0(980)}\sin\theta/\sqrt{2}$. Also note that the matrix
element $\la K^-(p_3)|(\bar s b)_{V-A}|B^-\ra \la
\pi^-(p_1)K^+(p_2)|(\bar d
 s)_{V-A}|0\ra$ has a similar expression as Eq. (\ref{eq:KpiBpi})
except for a sign difference
 \be
&& \la K^-(p_3)|(\bar s b)_{V-A}|B^-\ra \la
\pi^-(p_1)K^+(p_2)|(\bar d
 s)_{V-A}|0\ra \non\\
&=&
-F_1^{BK}(s_{12})F_1^{K\pi}(s_{12})\left[s_{23}-s_{13}-{(m_B^2-m_K^2)(m_K^2-m_\pi^2)
\over s_{12}}\right] \non
\\ && -F_0^{BK}(s_{12})F_0^{K\pi}(s_{12}){(m_B^2-m_K^2)(m_K^2-m_\pi^2)
\over s_{12}}.
 \en
As in Eq. (\ref{eq:F1Kpi}), the form factor $F_1^{K\pi}$ receives
a resonant contribution for the $K^*$ pole.

\begin{table}[t]
\caption{Same as Table \ref{tab:Kpipi} except for  $B^-\to
\pi^+\pi^-\pi^-$. The nonresonant background is used as an input to
fix the parameter $\alpha_{_{\rm NR}}$ defined in Eq.
(\ref{eq:ADalitz}). }
\begin{ruledtabular} \label{tab:Bpipipi}
\begin{tabular}{l l l }
 Decay mode~~ &  BaBar \cite{BaBarpipipi} & Theory  \\ \hline
 $\rho^0\pi^-$ & $8.8\pm1.0\pm0.6^{+0.1}_{-0.7}$ &  $7.7^{+0.0+1.7+0.3}_{-0.0-1.6-0.2}$ \\
 $f_0(980)\pi^-$ & $1.2\pm0.6\pm0.1\pm0.4<3.0$ & $0.39^{+0.00+0.01+0.03}_{-0.00-0.01-0.02}$ \\
NR  & $2.3\pm0.9\pm0.3\pm0.4<4.6$ & input \\ \hline
 Total  & $16.2\pm1.2\pm0.9$ & $12.0^{+1.1+2.0+0.4}_{-1.2-1.8-0.3}$ \\
\end{tabular}
\end{ruledtabular}
\end{table}

{\squeezetable
\begin{table}[t]
\caption{Same as Table \ref{tab:Kpipi} except for the decay $\ov
B^0\to \pi^+\pi^-\pi^0$.}
\begin{center} \label{tab:KpKmpim}
\begin{tabular}{|c| c c c c c| c |} \hline
Decay mode & ~~$\rho^+\pi^-$~~ & ~~$\rho^-\pi^+$~~ & ~~$\rho^0\pi^0$~~ & ~~$f_0(980)\pi^0$~~& ~~NR~~ & ~~Total~~ \\
\hline Theory & $8.5^{+0.0+1.1+0.2}_{-0.0-1.0-0.1}$ &
$15.5^{+0.0+4.0+0.3}_{-0.0-3.5-0.3}$ &
$1.0^{+0.0+0.3+0.0}_{-0.0-0.2-0.0}$ &
$0.010^{+0.000+0.003+0.000}_{-0.000-0.002-0.000}$ &
$0.05^{+0.02+0.01+0.00}_{-0.02-0.01-0.00}$
& $26.3^{+0.0+5.6+0.2}_{-0.0-5.0-0.2}$   \\
 \hline
\end{tabular}
\end{center}
\end{table}}

The nonresonant and various resonant contributions to $B^-\to
K^+K^-\pi^-$ are shown in Table \ref{tab:KpKmpim}. The predicted
total rate is consistent with upper limits set by BaBar and Belle.

\section{$B\to \pi\pi\pi$ decays}

The factorizable amplitudes of the tree-dominated decay $B^-\to
\pi^+\pi^-\pi^-$ and $\ov B^0\to \pi^+\pi^-\pi^0$ are given by
Eqs. (\ref{eq:3pi}) and (\ref{eq:pippimpi0}), respectively. We see
that the former is dominated by the $\rho^0$ pole, while the
latter receives $\rho^\pm$ and $\rho^0$ contributions. As a
consequence, the $\pi^+\pi^-\pi^0$ mode has a rate larger than
$\pi^+\pi^-\pi^-$ even though the former involves a $\pi^0$ in the
final state.

The $\pi^+\pi^-\pi^-$ mode receives nonresonant contributions
mostly from the $b\to u$ transition as the nonresonant
contribution in the matrix element $\la\pi^+\pi^-|\bar dd|0\ra$ is
suppressed by the smallness of penguin Wilson coefficients $a_6$
and $a_8$. Therefore, the measurement of the nonresonant
contribution in this decay can be used to constrain the
nonresonant parameter $\alpha_{_{\rm NR}}$ in Eq.
(\ref{eq:ADalitz})

\section{Direct $CP$ asymmetries}
Direct \CP asymmetries for various charmless three-body $B$ decays
are collected in Table \ref{tab:CP}. Mixing-induced and direct \CP
asymmetries  in $B^0\to K^+K^-K_{S,L}$ and $K_SK_SK_{S,L}$ decays
are already shown in Table \ref{tab:ASKKK}. It appears that direct
\CP violation is sizable in $K^+K^-K^-$ and $K^+K^-\pi^-$ modes.

The major uncertainty with direct \CP violation comes from the
strong phases which are needed to induce partial rate \CP
asymmetries. In this work, the strong phases arise from the
effective Wilson coefficients $a_i^p$ listed in (\ref{eq:ai}) and
from the Breit-Wigner formalism for resonances. Since direct \CP
violation in charmless two-body $B$ decays can be significantly
affected by final-state rescattering \cite{CCSfsi}, it is natural
to extend the study of final-state rescattering effects to the
case of three-body $B$ decays. We will leave this to a future
investigation.

\begin{table}[h]
\caption{Direct \CP asymmetries (in \%) for various charmless
three-body $B$ decays. For theoretical errors, see Table
\ref{tab:KpKmK0}. Experimental results are taken from \cite{HFAG}.}
\begin{ruledtabular} \label{tab:CP}
\begin{tabular}{l  cc r }
 Final state~~ & BaBar & Belle & Theory \\ \hline
 $K^+K^-K^-$ & $-2\pm3\pm2$ &  & $-10.4^{+1.7+0.9+0.9}_{-1.3-1.0-0.8}$
 \\
 $K^-K_SK_S$ & $-4\pm11\pm2$ & & $-3.9^{+0.0+0.6+0.3}_{-0.0-0.8-0.3}$ \\
 $K^+K^-\pi^-$ & $0\pm10\pm3$ & & $17.5^{+1.9+2.2+0.0}_{-3.8-3.4-0.2}$  \\
 $K^-\pi^+\pi^-$ & $-1.3\pm3.7\pm1.1$ & $4.9\pm2.6\pm2.0$ & $-3.3^{+0.7+0.4+0.3}_{-0.5-0.4-0.2}$ \\
 $K^-\pi^+\pi^0$ & & $7\pm11\pm1$ & $6.3^{+0.6+1.4+0.5}_{-0.7-1.4-0.5}$ \\
 $\ov K^0\pi^+\pi^-$ & & & $4.9^{+0.0+0.0+0.3}_{-0.0-0.1-0.4}$ \\
 $\ov K^0\pi^0\pi^0$ & $-23\pm52\pm13$ & $-17\pm24\pm6$ & $0.28^{+0.09+0.07+0.02}_{-0.06-0.06-0.02}$ \\
 $\ov K^0\pi^-\pi^0$ & & & $0.4^{+0.0+0.4+0.0}_{-0.0-0.4-0.0}$ \\
 $\pi^+\pi^-\pi^-$ &   $-1\pm8\pm3$ & & $4.4^{+0.8+1.2+0.0}_{-0.6-0.9-0.2}$ \\
 $\pi^+\pi^-\pi^0$ & & & $-3.0^{+0.1+0.2+0.3}_{-0.1-0.3-0.2}$ \\
\end{tabular}
\end{ruledtabular}
\end{table}

\section{Two-body $B\to VP$ and $B\to SP$ decays}

Thus far we have considered the branching ratio products $\B(B\to
Rh_1)\B(R\to h_2h_3)$ with the resonance $R$ being a vector meson
or a scalar meson. Using the experimental information on $\B(R\to
h_2h_3)$ \cite{PDG}
 \be
&& \B(K^{*0}\to K^+\pi^-)=\B(K^{*+}\to K^0\pi^+)=2\B(K^{*+}\to
K^+\pi^0)={2\over 3}, \non \\
&& \B(K_0^{*0}(1430)\to K^+\pi^-)=2\B(K_0^{*+}(1430)\to
 K^+\pi^0)={2\over 3}(0.93\pm0.10), \non \\ &&\B(\phi\to
 K^+K^-)=0.492\pm0.006\,.
 \en
one can extract the branching ratios of $B\to VP$ and $B\to SP$.
The results are summarized in Table \ref{tab:BR2body}.

\begin{table}[!]
\caption{Branching ratios of quasi-two-body decays $B\to VP$ and
$B\to SP$ obtained from the studies of three-body decays based on
the factorization approach. Unless specified, the experimental
results are obtained from the 3-body Dalitz plot analyses given in
previous Tables. Theoretical uncertainties have been added in
quadrature. QCD factorization (QCDF) predictions taken from
\cite{BN} for $VP$ modes and from \cite{CCY} for $SP$ channels are
shown here for comparison.}
\begin{ruledtabular} \label{tab:BR2body}
\begin{tabular}{l c c | c c}
 Decay mode~~ &  BaBar & Belle  & QCDF & This work  \\ \hline
 $\phi K^0$ & $8.4^{+1.5}_{-1.3}\pm0.5$  \footnotemark[1] &
 $9.0^{+2.2}_{-1.8}\pm0.7$  \footnotemark[2] & $4.1^{+0.4+1.7+1.8+10.6}_{-0.4-1.6-1.9-~3.0}$ & $5.3^{+1.0}_{-0.9}$ \\
 $\phi K^-$ & $8.4\pm0.7\pm0.7$ & $9.60\pm0.92^{+1.05}_{-0.84}$ &
 $4.5^{+0.5+1.8+1.9+11.8}_{-0.4-1.7-2.1-~3.3}$ & $5.9^{+1.1}_{-1.0}$ \\
 $\ov K^{*0}\pi^-$ & $13.5\pm1.2^{+0.8}_{-0.9}$ &
 $9.8\pm0.9^{+1.1}_{-1.2}$ & $3.6^{+0.4+1.5+1.2+7.7}_{-0.3-1.4-1.2-2.3}$ & $4.4^{+1.1}_{-1.0}$ \\
 $\ov K^{*0}\pi^0$ & $3.0\pm0.9\pm0.5$ & $<3.5$ & $0.7^{+0.1+0.5+0.3+2.6}_{-0.1-0.4-0.3-0.5}$
 & $1.5^{+0.5}_{-0.4}$ \\
 $K^{*-}\pi^+$ & $11.0\pm1.5\pm0.7$ & $8.4\pm1.1^{+0.9}_{-0.8}$ &
 $3.3^{+1.4+1.3+0.8+6.2}_{-1.2-1.2-0.8-1.6}$ & $3.1^{+0.9}_{-0.9}$ \\
 $K^{*-}\pi^0$ & $6.9\pm2.0\pm1.3$ \footnotemark[2] & & $3.3^{+1.1+1.0+0.6+4.4}_{-1.0-0.9-0.6-1.4}$ &
 $2.2^{+0.6}_{-0.5}$ \\
 $K^{*0}K^-$ & & &
 $0.30^{+0.11+0.12+0.09+0.57}_{-0.09-0.10-0.09-0.19}$ & $0.35^{+0.06}_{-0.06}$ \\
 $\rho^0K^-$ & $5.1\pm0.8^{+0.6}_{-0.9}$ &
 $3.89\pm0.47^{+0.43}_{-0.41}$ & $2.6^{+0.9+3.1+0.8+4.3}_{-0.9-1.4-0.6-1.2}$ & $1.3^{+1.9}_{-0.7}$  \\
 $\rho^0\ov K^0$ & $4.9\pm0.8\pm0.9$ & $6.1\pm1.0\pm1.1$
 & $4.6^{+0.5+4.0+0.7+6.1}_{-0.5-2.1-0.7-2.1}$ & $2.0^{+1.9}_{-0.9}$ \\
 $\rho^+K^-$ & $8.6\pm1.4\pm1.0$ & $15.1^{+3.4+2.4}_{-3.3-2.6}$ &
 $7.4^{+1.8+7.1+1.2+10.7}_{-1.9-3.6-1.1-~3.5}$  & $2.5^{+3.6}_{-1.4}$ \\
 $\rho^-\ov K^0$ & $8.0^{+1.4}_{-1.3}\pm0.5$ \footnotemark[2] & &
 $5.8^{+0.6+7.0+1.5+10.3}_{-0.6-3.3-1.3-~3.2}$ & $1.3^{+3.0}_{-0.9}$ \\
 $\rho^0\pi^-$ & $8.8\pm1.0^{+0.6}_{-0.9}$ &
 $8.0^{+2.3}_{-2.0}\pm0.7$  \footnotemark[2] & $11.9^{+6.3+3.6+2.5+1.3}_{-5.0-3.1-1.2-1.1}$  & $7.7^{+1.7}_{-1.6}$ \\
 $\rho^-\pi^+$ & & &
 $21.2^{+10.3+8.7+1.3+2.0}_{-~8.4-7.2-2.3-1.6}$ & $15.5^{+4.0}_{-3.5}$ \\
 $\rho^+\pi^-$ & & & $15.4^{+8.0+5.5+0.7+1.9}_{-6.4-4.7-1.3-1.3}$ & $8.5^{+1.1}_{-1.0}$  \\
 $\rho^0\pi^0$ & $1.4\pm0.6\pm0.3$ & $3.1^{+0.9+0.6}_{-0.8-0.8}$ &
 $0.4^{+0.2+0.2+0.9+0.5}_{-0.2-0.1-0.3-0.3}$ & $1.0^{+0.3}_{-0.2}$ \\
 $f_0(980)K^0;f_0\to \pi^+\pi^-$ & $5.5\pm0.7\pm0.6$ & $7.6\pm1.7^{+0.8}_{-0.9}$ &
 $6.7^{+0.1+2.1+2.3}_{-0.1-1.5-1.1}$ \footnotemark[3]   & $7.7^{+0.4}_{-0.7}$ \\
 $f_0(980)K^-;f_0\to \pi^+\pi^-$ & $9.3\pm1.0^{+0.6}_{-0.9}$ &
 $8.8\pm0.8^{+0.9}_{-1.8}$ & $7.8^{+0.2+2.3+2.7}_{-0.2-1.6-1.2}$ \footnotemark[3] & $7.7^{+0.4}_{-0.8}$ \\
 $f_0(980)K^0;f_0\to K^+K^-$ & $5.3\pm2.2$  &  & & $5.8^{+0.1}_{-0.5}$ \\
 $f_0(980)K^-;f_0\to K^+K^-$ & $6.5\pm2.5\pm1.6$ & $<2.9$ &  & $7.0^{+0.4}_{-0.7}$ \\
 $f_0(980)\pi^-;f_0\to \pi^+\pi^-$ & $<3.0$ & & $0.5^{+0.0+0.2+0.1}_{-0.0-0.1-0.0}$ \footnotemark[3]
 & $0.39^{+0.03}_{-0.02}$ \\
 $f_0(980)\pi^-;f_0\to K^+K^-$ & & & & $0.50^{+0.06}_{-0.04}$ \\
 $f_0(980)\pi^0;f_0\to \pi^+\pi^-$ & & &
 $0.02^{+0.01+0.02+0.04}_{-0.01-0.00-0.01}$ \footnotemark[3] & $0.010^{+0.003}_{-0.002}$ \\
 $\ov K^{*0}_0(1430)\pi^-$ & $36.6\pm1.8\pm4.7$ &
 $51.6\pm1.7^{+7.0}_{-7.4}$ & $11.0^{+10.3+7.5+49.9}_{-~6.0-3.5-10.1}$ & $16.9^{+5.2}_{-4.4}$ \\
 $\ov K^{*0}_0(1430)\pi^0$ & $12.7\pm2.4\pm4.4$ & $9.8\pm2.5\pm0.9$ &
 $6.4^{+5.4+2.2+26.1}_{-3.3-2.1-~5.7}$ & $6.8^{+2.3}_{-1.9}$ \\
 $K^{*-}_0(1430)\pi^+$ & $36.1\pm4.8\pm11.3$ & $49.7\pm3.8^{+4.0}_{-6.1}$
 & $11.3^{+9.4+3.7+45.8}_{-5.8-3.7-~9.9}$  & $16.2^{+4.7}_{-4.0}$ \\
 $K^{*-}_0(1430)\pi^0$ & & & $5.3^{+4.7+1.6+22.3}_{-2.8-1.7-~4.7}$
 &  $8.9^{+2.6}_{-2.2}$ \\
 $K^{*0}_0(1430)K^-$ & $<2.2$ \footnotemark[2] & & & $1.3^{+0.3}_{-0.3}$ \\
\end{tabular}
\end{ruledtabular}
 \footnotetext[1]{From the Dalitz plot analysis of $B^0\to K^+K^-K^0$
decay measured by BaBar (see Table \ref{tab:KpKmK0}), we obtain
$\B(B^0\to\phi K^0)=(6.2\pm0.9)\times 10^{-6}$. The experimental
value of BaBar cited in the Table is obtained from a direct
measurement of $B^0\to\phi K^0$.}
 \footnotetext[2]{not determined directly from the Dalitz plot analysis of three-body decays.}
 \footnotetext[3]{We have assumed
$\B(f_0(980)\to\pi^+\pi^-)=0.50$ for the QCDF calculation.}
\end{table}

Two remarks about the experimental branching ratios are in order:
(i) The BaBar results for the branching ratios of $\ov B^0\to
K^{*-}\pi^+,~\ov K^{*0}\pi^0,~K^{*-}_0(1430)\pi^+$ are inferred
from the three-body decays $\ov B^0\to \ov K^0\pi^+\pi^-$ (see
Table \ref{tab:Kmpippi0}) and Belle results are taken from $\ov
B^0\to K^-\pi^+\pi^0$ (see Table \ref{tab:K0pipi}). (ii) Branching
ratios of $\ov B^0\to \phi\ov K^0$ shown in Table
\ref{tab:BR2body} are not inferred from the Dalitz plot analysis
of $B\to KKK$ decays.

For comparison, the predictions of the QCD factorization approach
for $B\to VP$ \cite{BN} and $B\to SP$ \cite{CCY} are also
exhibited in Table \ref{tab:BR2body}. In order to compare theory
with experiment for $B\to f_0(980)K\to \pi^+\pi^- K$, we need an
input for $\B(f_0(980)\to \pi^+\pi^-)$. To do this, we shall use
the BES measurement \cite{BES}
 \be
 {\Gamma(f_0(980)\to \pi\pi)\over \Gamma(f_0(980)\to \pi\pi)+\Gamma(f_0(980)\to K\ov
K)}=0.75^{+0.11}_{-0.13}\,.
 \en
Assuming that the dominance of the $f_0(980)$ width by $\pi\pi$
and $K\ov K$ and applying isospin relation, we obtain
 \be
 \B(f_0(980)\to \pi^+\pi^-)=0.50^{+0.07}_{-0.09}\,, \qquad \B(f_0(980)\to
 K^+K^-)=0.125^{+0.018}_{-0.022}\,.
 \en
At first sight, it appears that the ratio defined by
 \be
 R\equiv{\B(f_0(980)\to K^+K^-)\over \B(f_0(980)\to
\pi^+\pi^-)}=0.25\pm0.06
 \en
is not consistent with the value of $0.69\pm0.32$ inferred from
the BaBar data (see Tables \ref{tab:KpKmKm} and \ref{tab:Kpipi})
 \be
 R={\Gamma(B^-\to f_0(980)K^-;f_0(980)\to
 K^+K^-)\over\Gamma(B^-\to f_0(980)K^-;
 f_0(980)\to\pi^+\pi^-)}={6.5\pm2.5\pm1.6\over
 9.3\pm1.0^{+0.6}_{-0.9}},
 \en
where we have applied the narrow width approximation Eq.
(\ref{eq:fact}).

The above-mentioned discrepancy can be resolved by noting that the
factorization relation Eq. (\ref{eq:fact}) for the resonant
three-body decay is applicable only when the two-body decays $B\to
RP$ and $R\to P_1P_2$ are kinematically allowed and the resonance
is narrow, the so-called narrow width approximation. However, as
the decay $f_0(980)\to K^+K^-$ is kinematically barely or even not
allowed, the off resonance peak effect of the intermediate
resonant state will become important. Therefore, it is necessary
to take into account the finite width effect of the $f_0(980)$
which has a width of order 40-100 MeV \cite{PDG}. In short, one
cannot determine the ratio $R$ by applying the narrow width
approximation to the three-body decays. That is, one should employ
the decays $B\to K\pi\pi$ rather than $B\to KKK$ to extract the
experimental branching ratio for $B\to f_0(980)K$ provided
$\B(f_0(980)\to\pi\pi)$ is available.

We now compare the present work for $B\to VP$ and $B\to SP$ with
the approach of QCD factorization \cite{BBNS,CCY}.  In this work,
our calculation of 3-body $B$ decays is similar to the simple
generalized factorization approach \cite{Ali,CCTY} by assuming a
set of universal and process independent effective Wilson
coefficients $a_i^p$ with $p=u,c$ in Eq. (\ref{eq:ai}).  In QCDF,
the calculation of $a_i^p$ is rather sophisticated. They are
basically the Wilson coefficients in conjunction with
short-distance nonfactorizable corrections such as vertex
corrections and hard spectator interactions. In general, they have
the expressions \cite{BBNS,BN}
 \be
 a_i^p(M_1M_2) &=& \left(c_i+{c_{i\pm1}\over N_c}\right)N_i(M_2)
  +{c_{i\pm1}\over N_c}\,{C_F\alpha_s\over
 4\pi}\Big[V_i(M_2)+{4\pi^2\over N_c}H_i(M_1M_2)\Big]+P_i^p(M_2),
 \en
where $i=1,\cdots,10$,  the upper (lower) signs apply when $i$ is
odd (even), $c_i$ are the Wilson coefficients,
$C_F=(N_c^2-1)/(2N_c)$ with $N_c=3$, $M_2$ is the emitted meson
and $M_1$ shares the same spectator quark with the $B$ meson. The
quantities $V_i(M_2)$ account for vertex corrections,
$H_i(M_1M_2)$ for hard spectator interactions with a hard gluon
exchange between the emitted meson and the spectator quark of the
$B$ meson and $P_i(M_2)$ for penguin contractions. Hence, the
effective Wilson coefficients $a_i^p(M_1M_2)$ depend on the nature
of $M_1$ and $M_2$; that is, they are process dependent. Moreover,
they depend on the order of the argument, namely,
$a_i^p(M_2M_1)\neq a_i^p(M_1M_2)$ in general. In the above
equation, $N_i(M_2)$ vanishes for $i=6,8$ and $M_2=V$, and equals
to unity otherwise. For three-body decays, in principle one should
also compute the vertex, gluon and hard spectator-interaction
corrections. Of course, these corrections for the three-body case
will be more complicated than the two-body decay one. One possible
improvement of the present work is to utilize the QCDF results for
the effective parameters $a_i^p(M_1M_2)$ in the vicinity of the
resonance region.

We next proceed to the comparison of numerical results. For $\phi
K,~K^*\pi$ and $K^*\ov K$ modes, the QCDF and the present work
have similar predictions. For the $\rho$ meson in the final
states, QCDF predicts slightly small $\rho K$ and too large
$\rho\pi$ compared to experiment. \footnote{Recall that the world
average of the branching ratio of $B^0\to\rho^\pm\pi^\mp$ is
$(24.0\pm2.5)\times 10^{-6}$ \cite{HFAG}, while QCDF predicts it
to be $\sim 36.6\times 10^{-6}$ \cite{BN}.}
In contrast, in the present work we obtain reasonable $\rho\pi$
but too small $\rho K$. This is ascribed to the form factor
$A_0^{B\rho}(0)=0.37\pm0.06$ employed in \cite{BN} that is too
large compared to ours $A_0^{B\rho}(0)=0.28\pm0.03$ (see Table
\ref{tab:FF}). Recall that the recent QCD sum rule calculation
also yields a smaller one $A_0^{B\rho}(0)=0.30^{+0.07}_{-0.03}$
\cite{Ball}.

For $B\to f_0(980)K$ and $B\to f_0(980)\pi$, QCDF \cite{CCY} and
this work are in agreement with experiment. The large rate of the
$f_0(980)K$ mode is ascribed to the large $f_0(980)$ decay
constant, $\bar f_{f_0(980)}\approx 460$ MeV at the
renormalization scale $\mu=2.1$ GeV \cite{CCY}. In contrast, the
predicted $\ov K^{*0}_0(1430)\pi^-$ and $K^{*-}_0(1430)\pi^+$ are
too small compared to the data. The fact that QCDF leads to too
small rates for $\phi K,~K^*\pi,~\rho K$ and $K_0^*(1430)\pi$ may
imply the importance of power corrections due to the non-vanishing
$\rho_A$ and $\rho_H$ parameters arising from weak annihilation
and hard spectator interactions, respectively, which are used to
parametrize  the endpoint divergences, or due to possible
final-state rescattering effects from charm intermediate states
\cite{CCSfsi}. However, this is beyond the scope of the present
work.

\section{Conclusions}

In this work, an exploratory study of charmless 3-body decays of
$B$ mesons is presented using a simple model based on the
framework of the factorization approach.  The 3-body decay process
consists of resonant contributions and the nonresonant signal.
Since factorization has not been proved for three-body $B$ decays,
we shall work in the phenomenological factorization model rather
than in the established theories such as QCD factorization. That
is, we start with the simple idea of factorization and see if it
works for three-body decays. Our main results are as follows:

\begin{itemize}

\item If heavy meson chiral perturbation theory (HMChPT) is
applied to the three-body matrix elements for $B\to P_1P_2$
transitions and assumed to be valid over the whole kinematic
region, then the predicted decay rates for nonresonant 3-body $B$
decays will be too large and even exceed the measured total rate.
This can be understood because chiral symmetry has been applied
beyond its region of validity. We assume the momentum dependence
of nonresonant amplitudes in the exponential form
$e^{-\alpha_{_{\rm NR}} p_B\cdot(p_i+p_j)}$ so that the HMChPT
results are recovered in the soft meson limit $p_i,~p_j\to 0$. The
parameter $\alpha_{_{\rm NR}}$ can be fixed from the
tree-dominated decay $B^-\to \pi^+\pi^-\pi^-$.

\item Besides the nonresonant contributions arising from $B\to
P_1P_2$ transitions, we have identified another large source of
the nonresonant background in the matrix elements of scalar
densities, e.g. $\la K\ov K|\bar ss|0\ra$ which can be constrained
from the $K_SK_SK_S$ (or $K^-K_SK_S$) mode in conjunction with the
mass spectrum in the decay $\ov B^0\to K^+K^-\ov K^0$ .

\item All $KKK$ modes are dominated by the nonresonant background.
The predicted branching ratios of $K^+K^-K_{S(L)}$, $K^+K^-K^-$
and $K^-K_SK_S$ modes are consistent with the data within the
theoretical and experimental errors.

\item Although the penguin-dominated $B^0\to K^+K^-K_{S}$ decay is
subject to a potentially significant tree pollution, its effective
$\sin 2\beta$ is very similar to that of the $K_SK_SK_S$ mode.
However, direct \CP asymmetry of the former, being of order
$-4\%$, is more prominent than the latter,

\item The role played by the unknown scalar resonance $X_0(1550)$
in the decay $B^-\to K^+K^-K^-$ should be clarified in order to
see if it behaves in the same way as in the $K^+K^-\ov K^0$ mode.

\item Applying SU(3) symmetry to relate the nonresonant component
in the matrix element $\la K\pi|\bar sq|0\ra$ to that in $\la K\ov
K|\bar ss|0\ra$, we found sizable nonresonant contributions in
$K^-\pi^+\pi^-$ and $\ov K^0\pi^+\pi^-$ modes, in agreement with
the Belle measurements but larger than the BaBar results. In
particular, the predicted nonresonant contribution in the
$K^-\pi^+\pi^0$ mode is consistent with the Belle limit and larger
than the BaBar's upper bound. It will be interesting to have a
refined measurement of the nonresonant contribution to this mode
to test our model.

\item  The $\pi^+\pi^-\pi^0$ mode is predicted to have a rate
larger than $\pi^+\pi^-\pi^-$ even though the former involves a
$\pi^0$ in the final state. This is because the latter is
dominated by the $\rho^0$ pole, while the former receives
$\rho^\pm$ and $\rho^0$ resonant contributions.

\item Among the 3-body decays we have studied, the decay $B^-\to
K^+K^-\pi^-$ dominated by $b\to u$ tree transition and $b\to d$
penguin transition has the smallest branching ratio of order
$4\times 10^{-6}$. It is consistent with the current bound set by
BaBar and Belle.

\item Decay rates and time-dependent \CP asymmetries in the decays
$K_S\pi^0\pi^0$, a purely $CP$-even state, and $K_S\pi^+\pi^-$, an
admixture of $CP$-even and $CP$-odd components, are studied. The
corresponding mixing-induced \CP violation is found to be of order
0.729 and 0.718, respectively.

\item Since the decay $f_0(980)\to K^+K^-$ is kinematically barely
or even not allowed, it is crucial to take into account the finite
width effect of the $f_0(980)$ when computing the decay $B\to
f_0(980)K\to KKK$. Consequently, one should employ the Dalitz plot
analysis of  $K\pi\pi$ mode to extract the experimental branching
ratio for $B\to f_0(980)K$ provided $\B(f_0(980)\to\pi\pi)$ is
available. The large rate of $B\to f_0(980)K$ is ascribed to the
large $f_0(980)$ decay constant, $\bar f_{f_0(980)}\approx 460$
MeV.

\item The intermediate vector meson contributions to 3-body decays
e.g. $\rho,~\phi,~K^*$ are identified through the vector current,
while the scalar meson resonances e.g.
$f_0(980),~X_0(1550),~K_0^*(1430)$ are mainly associated with the
scalar density. Their effects are described in terms of the
Breit-Wigner formalism.

\item Based on the factorization approach, we have computed the
resonant contributions to 3-body decays and determined the rates
for the quasi-two-body decays $B\to VP$ and $B\to SP$. The
predicted $\rho\pi,~f_0(980)K$ and $f_0(980)\pi$ rates are
consistent with experiment, while the calculated $\phi
K,~K^*\pi,~\rho K$ and $K_0^*(1430)\pi$ are too small compared to
the data.

\item Direct \CP asymmetries have been computed for the charmless
3-body $B$ decays. We found sizable direct \CP violation in
$K^+K^-K^-$ and $K^+K^-\pi^-$ modes.

\item In this exploratory work we use the phenomenological
factorization model rather than in the established theories based
on a heavy quark expansion. Consequently,  we don't have $1/m_b$
power corrections within this model. However, systematic errors due
to such model dependent assumptions may be sizable and are not
included in the error estimates that we give.

\end{itemize}

\vskip 2.0cm \noindent {\it Note added}: After the paper was
submitted for publication, BaBar (arXiv:0708.0367 [hep-ex]) has
reported the observation of the decay $B^+\to K^+K^-\pi^+$ with the
branching ratio $(5.0\pm0.5\pm0.5)\times 10^{-6}$. Our prediction
for this mode (see Table XIV) is consistent with experiment.

\vskip 2.0cm \acknowledgments   This research was supported in
part by the National Science Council of R.O.C. under Grant Nos.
NSC95-2112-M-001-013, NSC95-2112-M-033-013, and by the U.S. DOE
contract No. DE-AC02-98CH10886(BNL).

\newpage
\appendix

\section{Decay amplitudes of three-body $B$ decays}

In this appendix we list the factorizable amplitudes of the 3-body
decays $B\to KKK,KK\pi,K\pi\pi,\pi\pi\pi$. Under the factorization
hypothesis, the decay amplitudes are given by
 \be \label{eq:factamp}
 \la P_1P_2P_3|{\cal H}_{\rm eff}|\ov B\ra
 =\frac{G_F}{\sqrt2}\sum_{p=u,c}\lambda_p^{(r)} \la P_1P_2P_3|T_p|\ov B\ra,
 \en
where $\lambda_p^{(r)}\equiv V_{pb} V^*_{pr}$ with $r=d,s$. For
$KKK$ and $K\pi\pi$ modes, $r=s$ and for $KK\pi$ and $\pi\pi\pi$
channles, $r=d$. The Hamiltonian $T_p$ has the expression
\cite{BBNS}
 \be \label{eq:Tp}
 T_p&=&
 a_1 \delta_{pu} (\bar u b)_{V-A}\otimes(\bar s u)_{V-A}
 +a_2 \delta_{pu} (\bar s b)_{V-A}\otimes(\bar u u)_{V-A}
 +a_3(\bar s b)_{V-A}\otimes\sum_q(\bar q q)_{V-A}
 \non\\
 &&+a^p_4\sum_q(\bar q b)_{V-A}\otimes(\bar s q)_{V-A}
   +a_5(\bar s b)_{V-A}\otimes\sum_q(\bar q q)_{V+A}
       \non\\
 &&-2 a^p_6\sum_q(\bar q b)_{S-P}\otimes(\bar s q)_{S+P}
 +a_7(\bar s b)_{V-A}\otimes\sum_q\frac{3}{2} e_q (\bar q q)_{V+A}
 \non\\
 &&-2a^p_8\sum_q(\bar q b)_{S-P}\otimes\frac{3}{2} e_q
             (\bar s q)_{S+P}
 +a_9(\bar s b)_{V-A}\otimes\sum_q\frac{3}{2}e_q (\bar q q)_{V-A}\non\\
 &&+a^p_{10}\sum_q(\bar q b)_{V-A}\otimes\frac{3}{2}e_q(\bar s
 q)_{V-A},
 \en
with $(\bar q q')_{V\pm A}\equiv \bar q\gamma_\mu(1\pm\gamma_5)
q'$, $(\bar q q')_{S\pm P}\equiv\bar q(1\pm\gamma_5) q'$ and a
summation over $q=u,d,s$ being implied. For the effective Wilson
coefficients, we use
 \be \label{eq:ai}
 && a_1\approx0.99\pm0.037 i,\quad a_2\approx 0.19-0.11i, \quad a_3\approx -0.002+0.004i, \quad a_5\approx
 0.0054-0.005i,  \non \\
 && a_4^u\approx -0.03-0.02i, \quad a_4^c\approx
 -0.04-0.008i,\quad
 a_6^u\approx -0.06-0.02i, \quad a_6^c\approx -0.06-0.006i,
 \non\\
 &&a_7\approx 0.54\times 10^{-4} i,\quad a_8^u\approx (4.5-0.5i)\times
 10^{-4},\quad
 a_8^c\approx (4.4-0.3i)\times
 10^{-4},   \\
 && a_9\approx -0.010-0.0002i,\quad
 a_{10}^u \approx (-58.3+ 86.1 i)\times10^{-5},\quad
 a_{10}^c \approx (-60.3 + 88.8 i)\times10^{-5}, \non
 \en
for typical $a_i$ at the renormalization scale $\mu=m_b/2=2.1$~GeV
which we are working on.

Various three-body $B$ decay amplitudes are collected below.

\vskip 0.2 cm \noindent \underline{$B\to KKK$}

 \be  \label{eq:AKpKmK0}
 \la\overline K {}^0 K^+ K^-|T_p|\ov B^0\ra&=&
 \la K^+\ov K {}^0|(\bar u b)_{V-A}|\ov B {}^0\ra \la K^-|(\bar s u)_{V-A}|0\ra
 \left[a_1 \delta_{pu}+a^p_4+a_{10}^p-(a^p_6+a^p_8) r_\chi^K \right]
 \non\\
 &&+ \la K^+ K^-|(\bar d b)_{V-A}|\ov B {}^0\ra \la \ov K^0|(\bar s d)_{V-A}|0\ra
 \left(a^p_4-{1\over 2}a_{10}^p\right)
 \non\\
 &&+\la \ov K {}^0|(\bar s b)_{V-A}|\ov B {}^0\ra
                   \la K^+ K^-|(\bar u u)_{V-A}|0\ra
    (a_2\delta_{pu}+a_3+a_5+a_7+a_9)
                   \non\\
 &&+\la \ov K {}^0|(\bar s b)_{V-A}|\ov B {}^0\ra
                   \la K^+ K^-|(\bar d d)_{V-A}|0\ra
    \bigg[a_3+a_5-\frac{1}{2}(a_7+a_9)\bigg]
    \non\\
 &&+\la \ov K {}^0|(\bar s b)_{V-A}|\ov B {}^0\ra
                   \la K^+ K^-|(\bar s s)_{V-A}|0\ra
    \bigg[a_3+a^p_4+a_5-\frac{1}{2}(a_7+a_9+a^p_{10})\bigg]
    \non\\
 &&+\la \ov K {}^0|\bar s b|\ov B {}^0\ra
       \la K^+ K^-|\bar s s|0\ra
       (-2 a^p_6+a^p_8)
       \non\\
  &&+ \la K^+ K^-|\bar d(1-\gamma_5)b|\ov B {}^0\ra \la \ov K^0|\bar s(1+\gamma_5) d|0\ra
 \left(-2a^p_6+a_8^p\right)
   \non\\
  &&  +\la K^+ K^-\ov K {}^0|(\bar s d)_{V-A}|0\ra
     \la 0|(\bar d b)_{V-A}|\ov B {}^0\ra
       \bigg(a^p_4-\frac{1}{2} a^p_{10}\bigg)
       \non\\
 &&  + \la K^+ K^-\ov K {}^0|\bar s\gamma_5 d|0\ra
       \la 0|\bar d\gamma_5 b|\ov B {}^0\ra
       (-2a^p_6+a^p_8),
 \en
with $r_\chi^P={2 m_P^2\over m_b(\mu)(m_2+m_1)(\mu)}$.

 \be  \label{eq:AKpKmKm}
 \la K^+ K^- K^-|T_p|B^-\ra&=&
 \la K^+K^-|(\bar u b)_{V-A}|B^-\ra \la K^-|(\bar s u)_{V-A}|0\ra
 \left[a_1 \delta_{pu}+a^p_4+a_{10}^p-(a^p_6+a^p_8) r_\chi^K\right]
 \non\\
 &&+\la K^-|(\bar s b)_{V-A}| B^-\ra
                   \la K^+ K^-|(\bar u u)_{V-A}|0\ra
    (a_2\delta_{pu}+a_3+a_5+a_7+a_9)
                   \non\\
 &&+\la K^-|(\bar s b)_{V-A}| B^-\ra
                   \la K^+ K^-|(\bar d d)_{V-A}|0\ra
    \bigg[a_3+a_5-\frac{1}{2}(a_7+a_9)\bigg]
    \non\\
 &&+ \la K^-|(\bar s b)_{V-A}| B^-\ra
                   \la K^+ K^-|(\bar s s)_{V-A}|0\ra
    \bigg[a_3+a^p_4+a_5-\frac{1}{2}(a_7+a_9+a^p_{10})\bigg]
    \non\\
 &&+\la K^-|\bar s b|B^-\ra
       \la K^+ K^-|\bar s s|0\ra
       (-2 a^p_6+a^p_8)
       \non\\
  &&+ \la K^+ K^-|\bar u(1-\gamma_5)b|\ov B {}^0\ra \la K^-|\bar s(1+\gamma_5) u|0\ra
 \left(-2a^p_6+a_8^p\right)
   \non\\
  &&  +\la K^+ K^-K^-|(\bar s u)_{V-A}|0\ra
     \la 0|(\bar u b)_{V-A}|B^-\ra
       \bigg(a^p_4-\frac{1}{2} a^p_{10}\bigg)
       \non\\
 &&  + \la K^+ K^-K^-|\bar s\gamma_5 u|0\ra
       \la 0|\bar u\gamma_5 b|B^-\ra
       (-2a^p_6+a^p_8).
 \en
Since there are two identical $K^-$ mesons in this decay, one
should take into account the identical particle effects. For
example,
 \be
\la K^+K^-|(\bar u b)_{V-A}|B^-\ra \la K^-|(\bar s u)_{V-A}|0\ra
 &=& \la K^+(p_1)K^-(p_2)|(\bar u b)_{V-A}|B^-\ra \la K^-(p_3)|(\bar s u)_{V-A}|0\ra
 \non \\
 &+& \la K^+(p_1)K^-(p_3)|(\bar u b)_{V-A}|B^-\ra \la K^-(p_2)|(\bar s
 u)_{V-A}|0\ra,
 \non \\
 \en
and a factor of ${1\over 2}$ should be put in the decay rate.

 \be \label{eq:AK0K0K0}
 \la K^0 \ov K
 {}^0 \ov K {}^0|T_p|\ov B^0\ra &=&
 \la K^0\ov K {}^0|(\bar d b)_{V-A}|\ov B {}^0\ra \la \ov K {}^0|(\bar s d)_{V-A}|0\ra
 \Big(a^p_4-\frac{1}{2}a^p_{10}-(a^p_6-\frac{1}{2}a^p_8)
 r_\chi^K\Big)
 \non\\
            &&+\la \ov K {}^0 |(\bar s b)_{V-A}|\ov B {}^0\ra
       \la K^0 \ov K {}^0|(\bar d d)_{V-A}|0\ra
 \left[a_3+a_5-\frac{1}{2}(a_7+a_9)\right] \non\\
       &&+\la \ov K {}^0 |(\bar s b)_{V-A}|\ov B {}^0\ra
       \la K^0 \ov K {}^0|(\bar s s)_{V-A}|0\ra
 \left[a_3+a^p_4+a_5-\frac{1}{2}(a_7+a_9+a_{10}^p)\right] \non\\
  &&+\la \ov K {}^0 |\bar s b|\ov B {}^0\ra
       \la K^0 \ov K {}^0|\bar s s|0\ra
       (-2 a^p_6+a^p_8)
       \non\\
   &&  +   \la K^0 \ov K {}^0 \ov K {}^0|(\bar s d)_{V-A}|0\ra
      \la 0|(\bar db)_{V-A}|\ov B {}^0\ra
       \left(a_4^p-{1\over 2}(a_7+a_9+a_{10}^p)\right) \non \\
  &&  +   \la K^0 \ov K {}^0 \ov K {}^0|\bar s\gamma_5 d|0\ra
      \la 0|\bar d\gamma_5 b|\ov B {}^0\ra
       (-2a^p_6+a^p_8).
 \en
The second and third terms do not contribute to the purely
\CP-even decay $\ov B^0\to K_SK_SK_S$.

 \be \label{eq:AKmK0K0}
 \la K^- K
 {}^0 \ov K {}^0|T_p|B^-\ra &=&
   \la K^0\ov K^0|(\bar u b)_{V-A}|B^-\ra \la K^-|(\bar s u)_{V-A}|0\ra
 \left[a_1 \delta_{pu}+a^p_4+a_{10}^p-(a^p_6+a^p_8) r_\chi^K\right]
 \non\\
 && +\la K^0 K^-|(\bar d b)_{V-A}|B^-\ra \la \ov K {}^0|(\bar s d)_{V-A}|0\ra
 \Big(a^p_4-\frac{1}{2}a^p_{10}-(a^p_6-\frac{1}{2}a^p_8)
 r_\chi^K\Big)
 \non\\
       &&+\la K^- |(\bar s b)_{V-A}|B^-\ra
       \la K^0 \ov K {}^0|(\bar d d)_{V-A}|0\ra
 \left[a_3+a_5-\frac{1}{2}(a_7+a_9)\right] \non\\
       &&+\la K^- |(\bar s b)_{V-A}|B^-\ra
       \la K^0 \ov K {}^0|(\bar ss)_{V-A}|0\ra
 \left[a_3+a_4^p+a_5-\frac{1}{2}(a_7+a_9+a_{10}^p)\right] \non\\
  &&+\la K {}^- |\bar s b|B {}^-\ra
       \la K^0 \ov K {}^0|\bar s s|0\ra
       (-2 a^p_6+a^p_8)
       \non\\
   &&  +   \la K^- K {}^0 \ov K {}^0|(\bar s u)_{V-A}|0\ra
      \la 0|(\bar ub)_{V-A}|\ov B {}^0\ra
       (a_1\delta_{pu}+a_4^p+a_{10}^p) \non \\
  &&  +   \la K^- K {}^0 \ov K {}^0|\bar s\gamma_5u|0\ra
      \la 0|\bar u(1-\gamma_5) b|B {}^-\ra
       (2a^p_6+2a^p_8).
 \en
The third and fourth terms do not contribute to the decay $B^-\to
K^-K_SK_S$.

\vskip 0.2 cm \noindent \underline{$B\to KK\pi$}

 \be \label{eq:AKpKmpim}
 \la \pi^- K^+ K^-|T_p|B^-\ra &=&
 \la K^+ K^-|(\bar u b)_{V-A}|B^-\ra \la \pi^-|(\bar d u)_{V-A}|0\ra
 \left[a_1 \delta_{pu}+a^p_4+a_{10}^p-(a^p_6+a^p_8) r_\chi^\pi\right]
 \non\\
 &&+\la \pi^-|(\bar d b)_{V-A}|B ^-\ra
                   \la K^+ K^-|(\bar u u)_{V-A}|0\ra
    (a_2\delta_{pu}+a_3+a_5+a_7+a_9)
                   \non\\
  &&+\la \pi^-|\bar d b|B ^-\ra
                   \la K^+ K^-|\bar dd|0\ra
    (-2a_6^p+a_8^p)
                   \non\\
 &&+\la \pi^-|(\bar d b)_{V-A}|B ^-\ra
                   \la K^+ K^-|(\bar s s)_{V-A}|0\ra
    \bigg[a_3+a_5-\frac{1}{2}(a_7+a_9)\bigg]
    \non\\
&&+\la K^-|(\bar s b)_{V-A}|B^-\ra
       \la K^+\pi^-|(\bar d s)_{V-A}|0\ra
       (a^p_4-{1\over 2}a^p_{10})
       \non\\
 &&+\la K^-|\bar s b|B^-\ra
       \la K^+\pi^-|\bar d s|0\ra
       (-2 a^p_6+a^p_8)
       \non\\
  &&  +\la K^+ K^-\pi^-|(\bar d u)_{V-A}|0\ra
     \la 0|(\bar u b)_{V-A}|B^-\ra
       \bigg(a_1\delta_{pu}+a^p_4+a^p_{10}\bigg)
       \non\\
 &&  + \la K^+ K^-\pi^-|\bar d\gamma_5 u|0\ra
       \la 0|\bar u\gamma_5 b|B^-\ra
       (2a^p_6-a^p_8).
 \en

\vskip 0.2 cm \noindent \underline{$B\to K\pi\pi$}

 \be \label{eq:AKmpippim}
 \la K^- \pi^+ \pi^-|T_p|B^-\ra &=&
 \la \pi^+ \pi^-|(\bar u b)_{V-A}|B^-\ra \la K^-|(\bar s u)_{V-A}|0\ra
 \left[a_1 \delta_{pu}+a^p_4+a_{10}^p-(a^p_6+a^p_8) r_\chi^K\right]
 \non\\
&& + \la K^-|(\bar sb)_{V-A}|B^-\ra \la\pi^+\pi^-|(\bar
uu)_{V-A}|0\ra\left[a_2\delta_{pu}+{3\over 2}(a_7+a_9)\right] \non \\
&& +\la K^-|\bar sb|B^-\ra \la\pi^+\pi^-|\bar ss|0\ra
(-2a_6^p+a_8^p) \non \\
&& +\la\pi^-|(\bar db)_{V-A}|B^-\ra \la K^-\pi^+|(\bar
sd)_{V-A}|0\ra(a_4^p-{1\over 2}a_{10}^p) \non \\
&& +\la\pi^-|\bar db|B^-\ra \la K^-\pi^+|\bar
sd|0\ra(-2a_6^p+a_8^p) \non \\
&& +\la K^-\pi^+\pi^-|(\bar su)_{V-A}|0\ra \la0|(\bar
ub)_{V-A}|B^-\ra(a_1\delta_{pu}+a_4^p+a_{10}^p) \non \\
&& +\la K^-\pi^+\pi^-|\bar s\gamma_5u|0\ra \la0|\bar
u\gamma_5b|B^-\ra(2a_6^p+2a_8^p).
 \en

 \be \label{eq:AK0pippim}
 \la \ov K^0 \pi^+ \pi^-|T_p|\ov B^0\ra &=&
 \la \pi^+ \pi^-|(\bar d b)_{V-A}|\ov B^0\ra \la \ov K^0|(\bar s d)_{V-A}|0\ra
 \left[a^p_4-{1\over 2}a_{10}^p-(a^p_6-{1\over 2}a^p_8) r_\chi^K\right]
 \non\\
&& + \la \ov K^0|(\bar sb)_{V-A}|\ov B^0\ra \la\pi^+\pi^-|(\bar
uu)_{V-A}|0\ra\left[a_2\delta_{pu}+{3\over 2}(a_7+a_9)\right] \non \\
&& +\la \ov K^0|\bar sb|\ov B^0\ra \la\pi^+\pi^-|\bar ss|0\ra
(-2a_6^p+a_8^p) \non \\
&& +\la\pi^+|(\bar ub)_{V-A}|\ov B^0\ra \la \ov K^0\pi^-|(\bar
su)_{V-A}|0\ra(a_1+a_4^p+a_{10}^p) \non \\
&& +\la\pi^+|\bar ub|\ov B^0\ra \la \ov K^0\pi^-|\bar
su|0\ra(-2a_6^p-2a_8^p) \non \\
&& +\la \ov K^0\pi^+\pi^-|(\bar sd)_{V-A}|0\ra \la0|(\bar
db)_{V-A}|\ov B^0\ra(a_1\delta_{pu}+a_4^p+a_{10}^p) \non \\
&& +\la \ov K^0\pi^+\pi^-|\bar s(1+\gamma_5)d|0\ra \la0|\bar
d\gamma_5b|\ov B^0\ra(2a_6^p-a_8^p).
 \en

 \be \label{eq:AKmpippi0}
 \la K^- \pi^+ \pi^0|T_p|\ov B^0\ra &=&
 \la \pi^+ \pi^0|(\bar u b)_{V-A}|\ov B^0\ra \la K^-|(\bar s u)_{V-A}|0\ra
 \left[a_1\delta_{pu}+a^p_4+a_{10}^p-(a^p_6+a^p_8) r_\chi^K\right]
 \non \\
 && + \la K^-\pi^+|(\bar
sb)_{V-A}|\ov B^0\ra\la \pi^0|(\bar uu)_{V-A}|0\ra
\left[a_2\delta_{pu}
+{3\over 2}(-a_7+a_9)\right] \non \\
 &&  +\la \pi^+ |(\bar u b)_{V-A}|\ov B^0\ra \la K^-\pi^0|(\bar s
u)_{V-A}|0\ra \left[a_1\delta_{pu}+a^p_4+a_{10}^p\right]
 \non \\
&& +\la\pi^0|(\bar db)_{V-A}|\ov B^0\ra \la K^-\pi^+|(\bar
sd)_{V-A}|0\ra(a_4^p-{1\over 2}a_{10}^p) \non \\
 &&  +\la \pi^+ |\bar u b|\ov B^0\ra \la K^-\pi^0|\bar s
u|0\ra (-2a^p_6-2a_{8}^p)
 \non \\
&& +\la \pi^0|\bar db|\ov B^0\ra \la K^-\pi^+|\bar sd|0\ra
(-2a_6^p+a_8^p) \non \\
&& +\la K^-\pi^+\pi^0|(\bar sd)_{V-A}|0\ra \la0|(\bar
db)_{V-A}|\ov B^0\ra(a_4^p-{1\over 2}a_{10}^p) \non \\
&& +\la K^-\pi^+\pi^0|\bar s(1+\gamma_5)d|0\ra \la0|\bar
d\gamma_5b|\ov B^0\ra(2a_6^p-a_8^p).
 \en

 \be \label{eq:AK0pimpi0}
 \la \ov K^0 \pi^- \pi^0|T_p|B^-\ra &=&
 \la \pi^0 \pi^-|(\bar d b)_{V-A}|B^-\ra \la \ov K^0|(\bar s d)_{V-A}|0\ra
 \left[ a^p_4-{1\over 2}a_{10}^p-(a^p_6-{1\over 2}a^p_8) r_\chi^K\right]
 \non \\
 && + \la \ov K^0\pi^-|(\bar sb)_{V-A}|B^-\ra\la \pi^0|(\bar uu)_{V-A}|0\ra
\left[a_2\delta_{pu} +{3\over 2}(-a_7+a_9)\right] \non \\
  &&  +\la \pi^0|(\bar u b)_{V-A}|B^-\ra \la \ov K^0\pi^-|(\bar s
u)_{V-A}|0\ra \left[a_1\delta_{pu}+a_4^p+a_{10}^p\right]
 \non \\
 &&  +\la \pi^- |(\bar d b)_{V-A}|B^-\ra \la \ov K^0\pi^0|(\bar s
d)_{V-A}|0\ra \left[a^p_4-{1\over 2}a_{10}^p\right]
 \non \\
  &&  + \la \pi^0 |\bar u b|B^-\ra \la \ov K^0\pi^-|\bar s
 u|0\ra  (-2a^p_6-2a_{8}^p)
 \non \\
 &&  + \la \pi^- |\bar d b|B^-\ra \la \ov K^0\pi^0|\bar s
 d|0\ra  (-2a^p_6+a_{8}^p)
 \non \\
&& +\la \ov K^0\pi^-\pi^0|(\bar su)_{V-A}|0\ra \la0|(\bar
ub)_{V-A}|B^-\ra( a_1\delta_{pu}+a_4^p+a_{10}^p) \non \\
&& +\la \ov K^0\pi^-\pi^0|\bar s(1+\gamma_5)u|0\ra \la 0|\bar
u\gamma_5b|B^-\ra(2a_6^p+2a_8^p).
 \en

 \be \label{eq:AK0pi0pi0}
 \la \ov K^0 \pi^0 \pi^0|T_p|\ov B^0\ra &=&
 \la \pi^0 \pi^0|(\bar d b)_{V-A}|\ov B^0\ra \la \ov K^0|(\bar s d)_{V-A}|0\ra
 \left[ a^p_4-{1\over 2}a_{10}^p-(a^p_6-{1\over 2}a^p_8) r_\chi^K\right]
 \non \\
 && + \la \ov K^0\pi^0|(\bar sb)_{V-A}|\ov B^0\ra\la \pi^0|(\bar uu)_{V-A}|0\ra
\left[a_2\delta_{pu} +{3\over 2}(-a_7+a_9)\right] \non \\
 &&  +\la \pi^0 |(\bar d b)_{V-A}|\ov B^0\ra \la \ov K^0\pi^0|(\bar s
d)_{V-A}|0\ra \left[a^p_4-{1\over 2}a_{10}^p\right]
 \non \\
 && +\la\pi^0\pi^0|(\bar uu)_\vma|0\ra \la \ov K^0|(\bar
 sb)_\vma|\ov B^0\ra \left(a_2\delta_{pu}+2a_3+2a_5+{1\over
 2}(a_7+a_9)\right) \non \\
 &&  + \la \ov K^0\pi^0|\bar s
 d|0\ra \la \pi^0 |\bar d b|\ov B^0\ra  (-2a^p_6+a_{8}^p)
 \non \\
 && + \la\pi^0\pi^0|\bar ss|0\ra\la\ov K^0|\bar sb|\ov
 B^0\ra(-2a_6^p+a_8^p) \non \\
&& +\la \ov K^0\pi^0\pi^0|(\bar sd)_{V-A}|0\ra \la0|(\bar
db)_{V-A}|\ov B^0\ra(a_4^p-{1\over 2}a_{10}^p) \non \\
&& +\la \ov K^0\pi^0\pi^0|\bar s(1+\gamma_5)d|0\ra \la0|\bar
d\gamma_5b|\ov B^0\ra(2a_6^p-a_8^p).
 \en

\vskip 0.2 cm \noindent \underline{$B\to \pi\pi\pi$}

 \be \label{eq:3pi}
 \la \pi^- \pi^+ \pi^-|T_p|B^-\ra &=&
 \la \pi^+ \pi^-|(\bar u b)_{V-A}|B^-\ra \la \pi^-|(\bar d u)_{V-A}|0\ra
 \left[a_1 \delta_{pu}+a^p_4+a_{10}^p-(a^p_6+a^p_8) r_\chi^\pi\right]
 \non\\
&& + \la \pi^-|(\bar db)_{V-A}|B^-\ra \la\pi^+\pi^-|(\bar
uu)_{V-A}|0\ra\Big[a_2\delta_{pu}-a_4^p+{3\over 2}(a_7+a_9)+{1\over 2}a_{10}^p\Big] \non \\
&& +\la \pi^-|\bar db|B^-\ra \la\pi^+\pi^-|\bar dd|0\ra
(-2a_6^p+a_8^p) \non \\
&& +\la \pi^-\pi^+\pi^-|(\bar du)_{V-A}|0\ra \la0|(\bar
ub)_{V-A}|B^-\ra(a_1\delta_{pu}+a_4^p+a_{10}^p) \non \\
&& +\la \pi^-\pi^+\pi^-|\bar d(1+\gamma_5)u|0\ra \la0|\bar
u\gamma_5b|B^-\ra(2a_6^p+2a_8^p).
 \en

 \be \label{eq:pippimpi0}
 \la \pi^0 \pi^+ \pi^-|T_p|\ov B^0\ra &=&
 \la \pi^+ \pi^0|(\bar u b)_{V-A}|\ov B^0\ra \la \pi^-|(\bar d u)_{V-A}|0\ra
 \left[a_1 \delta_{pu}+a^p_4+a_{10}^p-(a^p_6+a^p_8) r_\chi^\pi\right]
 \non\\
  && + \la \pi^+ \pi^-|(\bar d b)_{V-A}|\ov B^0\ra \la \pi^0|(\bar u u)_{V-A}|0\ra
 \Big[a_2 \delta_{pu}-a_4^p+(a_6^p-{1\over 2}a_8^p)r_\chi^\pi \non \\
 &&~~~~~~~\qquad+{3\over 2}(a_7+a_9)+{1\over 2}a_{10}^p\Big]
 \non\\
  && + \la \pi^+ |(\bar u b)_{V-A}|\ov B^0\ra \la \pi^-\pi^0|(\bar d u)_{V-A}|0\ra
 \left[a_1 \delta_{pu}+a^p_4+a_{10}^p\right]
 \non\\
&& + \la \pi^0|(\bar db)_{V-A}|\ov B^0\ra \la\pi^+\pi^-|(\bar
uu)_{V-A}|0\ra\Big[a_2\delta_{pu}-a_4^p+{3\over 2}(a_7+a_9)+{1\over 2}a_{10}^p\Big] \non \\
&& +\la \pi^0|\bar db|\ov B^0\ra \la\pi^+\pi^-|\bar dd|0\ra
(-2a_6^p+a_8^p). \non \\
 \en

\section{Decay constants, form factors and others}

In this appendix we collect the numerical values of the decay
constants, form factors, CKM matrix elements and quark masses
needed for the calculations. We first discuss the decay constants
of the pseudoscalar meson $P$ and the scalar meson $S$ defined by
 \be \label{eq:decayc}
 && \la P(p)|\bar q_2\gamma_\mu\gamma_5 q_1|0\ra=-if_Pp_\mu, \qquad
 \la S(p)|\bar q_2\gamma_\mu q_1|0\ra=f_S p_\mu,
 \qquad \la S|\bar q_2q_1|0\ra=m_S\bar f_S,
 \en
and $\la V(p,\vp)|V_\mu|0\ra=f_Vm_V\vp_\mu^*$ for the vector
meson. For the scalar mesons, the vector decay constant $f_S$ and
the scale-dependent scalar decay constant $\bar f_S$ are related
by equations of motion
 \be \label{eq:EOM}
 \mu_Sf_S=\bar f_S, \qquad\quad{\rm with}~~\mu_S={m_S\over
 m_2(\mu)-m_1(\mu)},
 \en
where $m_{2}$ and $m_{1}$ are the running current quark masses.
The neutral scalar mesons $\sigma$, $f_0$ and $a_0^0$ cannot be
produced via the vector current owing to charge conjugation
invariance or conservation of vector current:
 \be
 f_{\sigma}=f_{f_0}=f_{a_0^0}=0.
 \en
However, the decay constant $\bar f_S$ is non-vanishing. In
\cite{CCY} we have applied the QCD sum rules to estimate this
quantity. In this work we folow \cite{CCY} to use
 \be
 \bar f_{f_0(980)}=460\,{\rm MeV}, \qquad
 \bar f_{K^*_0(1430)}=550\,{\rm MeV},
 \en
at $\mu=2.1$ GeV. As for the decay constants of vector mesons, we
use (in units of MeV).
 \be
 f_\rho=216, \qquad f_{K^*}=218, \qquad \bar f_{f_0(980)}=460,
\qquad \bar f_{K^*_0}=550.
 \en

Form factors for $B\to P,S$ transitions are defined by \cite{BSW}
 \be \label{eq:FF}
 \la P(p')|V_\mu|B(p)\ra &=& \left((p+p')_\mu-{m_B^2-m_P^2\over q^2}\,q_ \mu\right)
F_1^{BP}(q^2)+{m_B^2-m_P^2\over q^2}q_\mu\,F_0^{BP}(q^2), \non \\
\la S(p')|A_\mu|B(p)\ra &=&
-i\Bigg[\left((p+p')_\mu-{m_B^2-m_S^2\over q^2}\,q_ \mu\right)
F_1^{BS}(q^2)   +{m_B^2-m_S^2\over
q^2}q_\mu\,F_0^{BS}(q^2)\Bigg], \non \\
\la V(p',\vp)|V_\mu|B(p)\ra &=& {2\over
m_B+m_V}\,\epsilon_{\mu\nu\alpha \beta}\vp^{*\nu}p^\alpha p'^\beta
V(q^2),   \non \\
\la V(p',\vp)|A_\mu|B(p)\ra &=& i\Big[ (m_B+m_V)\vp^*_\mu
A_1^{BV}(q^2)-{\vp^*\cdot p\over m_B+m_V}\,(p+p')_\mu
A_2^{BV}(q^2) \non  \\ && -2m_V\,{\vp^*\cdot p\over
q^2}\,q_\mu\big[A_3^{BV}(q^2)-A_0^{BV}(q^2)\big]\Big],
 \en where
$q=p-p'$, $F_1(0)=F_0(0)$, $A_3(0)=A_0(0)$, and \be
A_3(q^2)=\,{m_P+m_V\over 2m_V}\,A_1(q^2)-{m_P-m_V\over
2m_V}\,A_2(q^2),
 \en
where $P_\mu=(p+p')_\mu$, $q_\mu=(p-p')_\mu$.  As shown in
\cite{CCH}, a factor of $(-i)$ is needed in $B\to S$ transition in
order for the $B\to S$ form factors to be positive. This also can
be checked from heavy quark symmetry \cite{CCH}.

Various form factors for $B\to S$ transitions have been evaluated
in the relativistic covariant light-front quark model \cite{CCH}.
In this model form factors are first calculated in the spacelike
region and their momentum dependence is fitted to a 3-parameter
form
  \be \label{eq:FFpara}
 F(q^2)=\,{F(0)\over 1-a(q^2/m_{B}^2)+b(q^2/m_{B}^2)^2}\,.
 \en
The parameters $a$, $b$ and $F(0)$ are first determined in the
spacelike region. This parametrization is then analytically
continued to the timelike region to determine the physical form
factors at $q^2\geq 0$.  The results relevant for our purposes are
summarized in Table \ref{tab:FF}.  In practical calculations, we
shall assign the form factor error to be $0.03$. For example,
$F_{0,1}^{BK}(0)=0.35\pm0.03$.

The form factor for $B$ to $f_0(980)$ is of order 0.25 at $q^2=0$
\cite{CCY}. In the $q\bar q$ model for the $f_0(980)$,
$F^{Bf_0^u}=F^{Bf_0}\sin\theta/\sqrt{2}$.

For the heavy-flavor independent strong coupling $g$ in HMChPT, we
use $|g|=0.59\pm0.01\pm0.07$ as extracted from the CLEO
measurement of the $D^{*+}$ decay width \cite{CLEOg}. The sign is
fixed to be negative in the quark model \cite{Yan}.


\begin{table}[h]
\caption{Form factors of $B\to\pi,K,K_0^*(1430),\rho$ transitions
obtained in the covariant light-front model \cite{CCH}. }
\label{tab:FF}
\begin{ruledtabular}
\begin{tabular}{| l c c c c | l c c c c |}
~~~$F$~~~~~
    & $F(0)$~~~~~
    & $F(q^2_{\rm max})$~~~~
    &$a$~~~~~
    & $b$~~~~~~
& ~ $F$~~~~~~~~
    & $F(0)$~~~~~
    & $F(q^2_{\rm max})$~~~~~
    & $a$~~~~~
    & $b$~~~~~~
 \\
    \hline
$F^{B\pi}_1$
    & $0.25$
    & $1.16$
    & 1.73
    & 0.95
& $F^{B\pi}_0$
    & 0.25
    & 0.86
    & 0.84
    & $0.10$
    \\
$F^{BK}_1$
    & $0.35$
    & $2.17$
    & 1.58
    & 0.68
& $F^{BK}_0$
    & 0.35
    & 0.80
    & 0.71
    & $0.04$
    \\
$F^{BK^*_0}_1$
    & $0.26$
    & $0.70$
    & 1.52
    & 0.64
&$F^{BK^*_0}_0$
    & 0.26
    & 0.33
    & 0.44
    & 0.05
    \\
$V^{B\rho}$
    & $0.27$
    & $0.79$
    & 1.84
    & 1.28
&$A^{B\rho}_0$
    & 0.28
    & 0.76
    & 1.73
    & 1.20
    \\
$A^{B\rho}_1$
    & 0.22
    & 0.53
    & 0.95
    & 0.21
&$A^{B\rho}_2$
    & $0.20$
    & $0.57$
    & 1.65
    & 1.05
 \\
$V^{BK^*}$
    & $0.31$
    & $0.96$
    & 1.79
    & 1.18
&$A^{BK^*}_0$
    & 0.31
    & 0.87
    & 1.68
    & 1.08
    \\
$A^{BK^*}_1$
    & 0.26
    & 0.58
    & 0.93
    & 0.19
&$A^{BK^*}_2$
    & $0.24$
    & $0.70$
    & 1.63
    & 0.98
    \\
\end{tabular}
\end{ruledtabular}
\end{table}


For the CKM matrix elements, we use the Wolfenstein parameters
$A=0.806$, $\lambda=0.22717$, $\bar \rho=0.195$ and $\bar
\eta=0.326$ \cite{CKMfitter}. The corresponding CKM angles are
$(\sin2\beta)_{CKM}=0.695^{+0.018}_{-0.016}$ and
$\gamma=(59\pm7)^\circ$ \cite{CKMfitter}. For the running quark
masses we shall use
 \be
 && m_b(m_b)=4.2\,{\rm GeV}, \qquad~~~ m_b(2.1\,{\rm GeV})=4.95\,{\rm
 GeV}, \qquad m_b(1\,{\rm GeV})=6.89\,{\rm
 GeV}, \non \\
 && m_c(m_b)=1.3\,{\rm GeV}, \qquad~~~ m_c(2.1\,{\rm GeV})=1.51\,{\rm
 GeV}, \non \\
 && m_s(2.1\,{\rm GeV})=90\,{\rm MeV}, \quad m_s(1\,{\rm GeV})=119\,{\rm
 MeV}, \non\\
 && m_d(1\,{\rm GeV})=6.3\,{\rm  MeV}, \quad~ m_u(1\,{\rm GeV})=3.5\,{\rm
 MeV}.
 \en
The uncertainty of the strange quark mass is specified as
$m_s(2.1\,{\rm GeV})=90\pm20$ MeV.


\end{document}